\documentclass[aps,prd,twocolumn,floatfix,nofootinbib,superscriptaddress,tightenlines]{revtex4}
\usepackage[dvipsnames]{xcolor}
\usepackage{amsmath}
\usepackage[utf8]{inputenc}
\usepackage{dcolumn}
\usepackage{lipsum}
\usepackage{amssymb}
\usepackage{soul}
\usepackage{url}
\usepackage{epsfig}
\usepackage{graphicx}
\usepackage{amsmath}
\usepackage{bm}
\usepackage{setspace}
\usepackage{appendix}
\usepackage{lscape}
\usepackage{amsthm}
\usepackage{bbold}
\usepackage{dcolumn}
\usepackage{epsfig}
\usepackage{graphics}
\usepackage{graphicx}
\usepackage{longtable}
\usepackage[utf8]{inputenc}

\usepackage{slashed}
\usepackage{bm}
\usepackage{xspace}
\usepackage{cancel}
\usepackage{float}
\usepackage{multirow}
\definecolor{darkgreen}{rgb}{0,0.5,0}
\definecolor{purple}{rgb}{0.5,0,0.5}
\definecolor{nblue}{rgb}{0.0,0.0,0.50}
\definecolor{scarlet}{rgb}{1.0,0.2,0}
\definecolor{darkmagenta}{rgb}{0.55, 0.0, 0.55}
\definecolor{darkolivegreen}{rgb}{0.33, 0.42, 0.18}
\definecolor{darkcandyapplered}{rgb}{0.64, 0.0, 0.0}

\usepackage[colorlinks=true, pdfstartview=FitV, linkcolor=purple, citecolor= purple, urlcolor=blue]{hyperref}

\newcommand{\be}{\begin{equation}}
\newcommand{\tu}{\textcolor{red}{u}}

\newcommand{\Jpsi}{\textcolor{blue}{J/\Psi}}

\newcommand{\td}{\textcolor{darkcandyapplered}{d}}
\newcommand{\tb}{\textcolor{blue}{b}}
\newcommand{\tc}{\textcolor{darkmagenta}{c}}
\newcommand{\ts}{\textcolor{darkgreen}{s}}
\newcommand{\ee}{\end{equation}}
\newcommand{\bea}{\begin{eqnarray}}
\newcommand{\eea}{\end{eqnarray}}
\newcommand{\beas}{\begin{eqnarray*}}
\newcommand{\eeas}{\end{eqnarray*}}

\begin{document}
\title{Spectroscopy of $\rho$-meson in symmetric nuclear medium}
\author{Anshu Gautam}
\email{gautamanshu681@gmail.com}
\author{Tanisha}
\email{tanisha220902@gmail.com}
\author{Satyajit Puhan}
\email{puhansatyajit@gmail.com}
\author{Arvind Kumar}
\email{kumara@nitj.ac.in}
  \author{Harleen Dahiya}
\email{dahiyah@nitj.ac.in}
\affiliation{Computational High Energy Physics Lab, Department of Physics, Dr. B.R. Ambedkar National
	Institute of Technology, Jalandhar, 144008, India}

\begin{abstract}
In this work, we investigate the behavior of the light vector \(\rho\) meson in the presence of a symmetric nuclear medium at zero temperature. We calculate the mass and decay constant of the $\rho$-meson as well as the leading twist distribution amplitudes (DAs) in the light-front quark model in vacuum, which are further investigated at different baryonic densities. We also predict the Mellin moments of the DAs and decay width of the $\rho^0 \to e^+ e^-$ process in both vacuum and medium. The evolution of DAs is carried out by the leading order (LO) Efremov-Radyushkin-Brodsky-Lepage method and compared with available predictions. For better understanding of medium effects on $\rho$-meson, we have also predicted the in-medium charge ($G_C(Q^2)$), magnetic ($G_M(Q^2)$), and quadrupole ($G_Q(Q^2)$) form factors. The in-medium charge radii, magnetic moment, and quadrupole moment have also been predicted in this work. We have found that the nuclear medium induces appreciable modifications on the mass, weak decay constant, decay width, and distribution amplitudes of the \(\rho\) meson. However, the charge radii, magnetic moment, and quadrupole moment are observed to exhibit weaker sensitivity to changes in baryonic density.
\end{abstract}


\maketitle

\section{Introduction}
Understanding the internal structure of hadrons is always a challenging task due to quark confinement at low energy, the non-perturbative quantum chromodynamics (QCD) limit. The study of strong interaction among the constituent quarks, gluons, and sea-quarks that make up the hadron is still a puzzle in understanding the asymptotic freedom and color confinement at different energy scales. It is also more complex to study the behavior of these constituents in the presence of a dense nuclear medium, which plays a crucial role in studying the chiral symmetry breaking at low energy scale and its restoration at higher densities of Nambu-Goldstone bosons (NGBs) \cite{Hayano:2008vn, Brown:1995qt, Brown:2001nh}. Modifications in the properties of hadrons and their constituents in the presence of nuclear medium have also been explored experimentally and theoretically after verification of the suppression in the structure function of nucleons in the presence of a nucleus to that of the free space in the region $0.3 <x<0.7$ ($x$ is the momentum fraction carried by the constituents) by the European Muon Collaboration (EMC) \cite{EuropeanMuon:1983wih}. Similar observations have also been verified in deep inelastic scattering (DIS) experiments by Stanford Linear Accelerator Center (SLAC) in the presence of light and heavy nuclei \cite{Arnold:1983mw}. Thus, the modification in the mass spectra, decay constant, charge radii, parton distribution functions (PDFs), distribution amplitudes (DAs), and other properties is expected in the presence of nuclear medium \cite{Eletsky:1996jg}. Experimentally, a reduction in the pion decay constant has been observed in the pion-nucleus scattering \cite{Friedman:2004jh} and the deep-bound pionic atom experiments \cite{Suzuki:2002ae}. The partial restoration of chiral symmetry breaking has also been experimentally confirmed in deeply bound pionic atoms \cite{Suzuki:2002ae}, di-pion
production in hadron-nucleus and photon-nucleus reactions \cite{CHAOS:1996nql,CHAOS:2004rhl}, and low-energy pion-nucleus scattering \cite{Friedman:2004jh}. So understanding the nuclear modifications is an important task for both experimentalists and theorists.

\par 
The $\rho$ meson serves as an ideal probe to understand the medium modifications of vector mesons due to its strong coupling to the $\pi \pi$ channel. Significant changes in the decay constant and decay width are expected within the medium due to the \(\rho\) meson's short lifetime of $1.3$ fm/c. The first indication of possible modification in the  $\rho$ meson mass due to a medium was observed in CERES \cite{CERES:1995vll} and HELIOS/3 \cite{Mazzoni:1994rb} collaborations at CERN. Subsequent studies by the upgraded CERES \cite{CERES:2006wcq}, NA60 \cite{NA60:2006ymb}, and CLAS \cite{CLAS:2007dll} collaborations further explored these effects, revealing notable changes in the decay width, mass spectra, and spectral function of the $\rho$ meson in the nuclear medium. Theoretical investigations of these in-medium properties have also been explored using QCD sum rules \cite{Ruppert:2005id, Kwon:2008vq, Asakawa:1993pq}, nuclear photoproduction \cite{Riek:2010gz}, quark-meson coupling (QMC) model \cite{deMelo:2018hfw}, dense hadronic matter \cite{Asakawa:1992ht}, dilepton production \cite{Chanfray:1993ue}, etc. 

\par In this work, we investigate the in-medium properties of the $\rho$ meson in symmetric nuclear matter using a hybrid approach that combines the light-front quark model (LFQM) and the chiral SU(3) quark mean-field model (CQMF). The vacuum properties of the $\rho$ meson are evaluated within the LFQM, while its in-medium behavior is analyzed using the CQMF by incorporating medium-modified input parameters. We have studied the meson mass, decay constant, DAs, decay width, life-time, Mellin moment, and form factors in vacuum as well as at different baryonic densities. This hybrid approach provides a benchmark for studying the in-medium properties of hadrons and their interactions in light-front dynamics (LFD). Similar hybrid approaches have been utilized to investigate the properties of pions and kaons in the quark meson coupling model (QMC), combined with the Nambu–Jona-Lasinio model (NJL) \cite{Hutauruk:2018qku, Hutauruk:2019ipp}, the LFQM with the QMC model \cite{Arifi:2024tix, Arifi:2023jfe}, and the light-cone quark model (LCQM) with the CQMF \cite{Puhan:2024xdq}. However, the in-medium properties of the vector $\rho$ meson have only been studied using the NJL model alongside the light-front constituent quark model (LFCQM) \cite{deMelo:2018hfw}. These hybrid models show qualitative in-medium results for both quark and meson properties. This highlights the need for more research on the properties of vector mesons in a nuclear medium.

\par The vacuum properties of the $\rho$ meson have been calculated using the LFQM. LFQM is a non-perturbative model based on light-front dynamics at a low energy scale. It serves as an effective framework for studying high-energy processes and parton distributions, as it simplifies the vacuum structure, ensures boost invariance, and facilitates the application of factorization theorems. The meson mass of LFQM is calculated by solving the QCD-motivated effective Hamiltonian $H_{q\bar q}=M_{0}+V_{q\bar q}$, where $V_{q\bar q}$ represents the quark-antiquark interaction potential. The mass spectrum of the meson is then determined from the eigenvalue equation $H_{q\bar q}|\Psi\rangle =(M_0+V_{q\bar q})|\Psi\rangle$. Various studies \cite{Arifi:2022pal, Dhiman:2019ddr, Choi:2015ywa, Choi:1997iq} have investigated the vacuum masses of vector mesons within the LFQM framework by employing different forms of the interaction potential $V_{q\bar q}$. After the indication of a change in decay width due to the presence of a nuclear medium by CLAS \cite{CLAS:2007dll}, understanding these effects theoretically has become a significant challenge. While the LFQM effectively describes the vacuum properties of mesons, it does not inherently account for the impact of the surrounding medium. To address this limitation and to extend the study beyond vacuum conditions, in-medium effects are incorporated through constituent quark masses determined by the CQMF model. 
\par The CQMF model is a theoretical framework that incorporates both the spontaneous and explicit breaking of chiral symmetry, which are essential features of low-energy QCD. Within this framework, quarks interact through the exchange of scalar (\(\sigma, \zeta, \delta\)) and vector (\(\omega, \rho\)) meson fields, while being contained within baryons via a confining potential. The model represents the interactions between quarks and mesons using a mean-field approximation, treating meson fields as classical entities. Consequently, only the scalar and vector fields play a role in the quark-meson interaction Lagrangian. The CQMF model has been utilized to explore the characteristics of nuclear and strange hadronic matter \cite{Wang:2001jw, Wang:2001hw}.  It has been used to theoretically compute the magnetic moment of octet and decuplet baryons in nuclear and strange hadronic medium \cite{Singh:2016hiw, Singh:2020nwp, Singh:2018kwq}. Furthermore, the combined approaches of the CQMF model and LCQM have been used to explore the in-medium properties such as decay constant, DAs, PDFs, generalized parton distributions (GPDs), as well as transverse momentum-dependent parton distributions (TMDs) of kaons and pions \cite{Puhan:2025ibn, Singh:2024lra, Puhan:2024xdq, Kaur:2024wze}.

\par The outline of the paper is as follows: In Section \ref{Sec2}, we describe the methodology, which is divided into two parts. In Subsection \ref{Sec2.1}, we provide an overview of the chiral quark mean field (CQMF) model used to obtain the in-medium quark masses, while in Subsection \ref{Sec2.2}, we outline the light-front quark model (LFQM) adopted in this work. In Section \ref{Sec3}, we present the results for the in-medium \(\rho\) meson mass, and in Section \ref{Sec4}, we discuss the calculation of the decay width and weak decay constant. The medium modification of the DAs is analyzed in Section \ref{Sec5}, followed by the evaluation of the in-medium form factors in Section \ref{Sec6}. Finally, in Section \ref{Sec7}, we summarize the findings of our study.

\section{Methodology} 
\label{Sec2}
Understanding hadronic characteristics in a dense medium is crucial for understanding the non-perturbative phase of  QCD, especially with regard to hadron spectral function modification and partial restoration of chiral symmetry. The light vector \(\rho\) meson is an important meson among others because of its short lifetime and strong coupling to dilepton channels, which makes it a sensitive probe of the dynamics in the medium. Its mass, decay width, and spectral shape changes inside thermal or nuclear matter have been well studied experimentally and theoretically, particularly in heavy-ion collisions.
 
This section discusses the CQMF model that uses medium-modified quark masses as input in the LFQM to study the properties of \(\rho\) meson within the nuclear medium. 

\subsection{Chiral SU(3) quark mean field model}
\label{Sec2.1}
The CQMF model respects the chiral symmetry and its spontaneous breaking \cite{Coleman:1969sm, Weinberg:1968de} as well as incorporates the effects of broken scale invariance \cite{Papazoglou:1998vr, Mishra:2003se, Mishra:2003tr}, to represent interactions between mesons and quarks and among mesons at finite baryon density. In this model, quarks are bound within baryons through a confining potential. The interaction of quarks with scalar-isoscalar fields \(\sigma\) and \(\zeta\), the scalar-isovector field \(\delta\), and the vector-isoscalar \(\omega\), along with the vector-isovector field \(\rho\), leads to significant modifications in the internal structure of baryons. Furthermore, to incorporate the trace anomaly of QCD, the model includes the scalar dilaton field \(\chi\) \cite{Papazoglou:1998vr, Kumar:2018ujk}.
\par The thermodynamic potential for isospin asymmetric nuclear matter at finite temperature and density is written as
\begin{eqnarray}
\Omega &=& - \frac{k_B T}{(2\pi)^3} \sum_{i=p/n} \gamma_i \int_0^\infty d^3k \left[ 
\ln \left( 1 + e^{-\left(\frac{E^*_i(k) - \nu^*_i}{k_B T}\right)} \right) \right. \nonumber \\
&& \left. + \ln \left( 1 + e^{-\left(\frac{E^*_i(k) + \nu^*_i}{k_B T}\right)} \right) 
\right] - \mathcal{L}_M - \mathcal{V}_{\text{vac}}.
\label{eqn:1}
\end{eqnarray}
Here, the summation is taken over the nucleons in the medium denoted by \(i=p/n\), and \(\gamma_i=2\) accounts for the spin degeneracy. The effective  energy $E_i^*(k)$  and the effective chemical potential $\nu_i^*$ of the baryon are given by  
\begin{eqnarray}
E_i^* &=& \sqrt{{M_i^*}^2 + k^2}, \\
\nu^*_i &=& \nu_i - g_\omega^i \, \omega - g_\rho^i \, I^{3i} \rho.
\end{eqnarray}
In the above, \(M_i^*\) denotes the effective mass of the baryon, \(\nu_i\) represents the free chemical potential, while \(g_\omega^i\) and \(g_\rho^i\) are the coupling constants of the vector fields \(\omega\) and \(\rho\), respectively. The third component of isospin for the nucleons is given by \(I^{3p}=-I^{3n}=1/2\). The thermodynamic potential defined by Eq.~\eqref{eqn:1} is minimized with respect to the scalar, vector, and dilaton field through
\begin{eqnarray}
\frac{\partial \Omega}{\partial \sigma}=
\frac{\partial \Omega}{\partial \zeta}=
\frac{\partial \Omega}{\partial \delta} =
\frac{\partial \Omega}{\partial \chi} =
\frac{\partial \Omega}{\partial \omega} = 
\frac{\partial \Omega}{\partial \rho} = 0.
\end{eqnarray}
This results in a set of nonlinear equations that are solved for various baryon density and temperature values. The interactions within the CQMF model are described by the effective Lagrangian density, which is given by \cite{Wang:2001jw}
\begin{eqnarray}
\label{eqn:2}
\mathcal{L}_{\text{eff}} &=& \mathcal{L}_{q0} + \mathcal{L}_{qm} + \mathcal{L}_{\Sigma \Sigma} + \mathcal{L}_{VV}+ \mathcal{L}_{\chi SB} \nonumber \\&& + \mathcal{L}_{\Delta m} + \mathcal{L}_{c}.
\end{eqnarray}
In the above, $ \mathcal{L}_{q0}$ denotes the kinetic term and $\mathcal{L}_{qm}$ represents the quark-meson interaction term, including quark interactions with spin-0 and spin-1 mesons \cite{Wang:2001jw, Singh:2016hiw, Wang:2001hw}. The expressions for \(\mathcal{L}_{q0}\) and \(\mathcal{L}_{qm}\) are written as
\begin{eqnarray}
\mathcal{L}_{q0} &=& \bar{\Psi} \, i \gamma^\mu \partial_\mu \Psi,
\end{eqnarray}
and 
\begin{eqnarray}
\mathcal{L}_{qm} &=& g_s \left(\bar{\Psi}_L M \Psi_R + \bar{\Psi}_R M^\dagger \Psi_L\right) \nonumber \\
& & {}- g_v \left(\bar{\Psi}_L \gamma^\mu l_\mu \Psi_L + \bar{\Psi}_R \gamma^\mu r_\mu \Psi_R\right) \nonumber \\
&=& \frac{g_s}{\sqrt{2}} \, \bar{\Psi} \left( \sum_{a=0}^8 s_a \lambda_a + i \gamma^5 \sum_{a=0}^8 p_a \lambda_a \right) \Psi \nonumber \\
& & {}- \frac{g_v}{2\sqrt{2}} \, \bar{\Psi} \gamma^\mu\left(  \sum_{a=0}^8 v_a^\mu \lambda_a 
- \gamma^5 \sum_{a=0}^8 a^\mu_a \lambda_a \right) \Psi,
\end{eqnarray}
respectively. Here, \( \Psi = \begin{pmatrix} u \\ d \end{pmatrix} \)
and parameters $g_s$ and $g_v$ are scalar and vector coupling constants. In Eq.\eqref{eqn:2}, \(\mathcal{L}_{\Sigma \Sigma}\) represent the chiral invariant self-interaction term for scalar mesons and is expressed as 
\begin{align}
\mathcal{L}_{\Sigma\Sigma} =\ & 
-\frac{1}{2} k_0 \chi^2 \left(\sigma^2 + \zeta^2 + \delta^2\right) 
+ k_1 \left(\sigma^2 + \zeta^2 + \delta^2\right)^2 \nonumber \\
& + k_2 \left(\frac{\sigma^4}{2} + \frac{\delta^4}{2} + 3\sigma^2\delta^2 + \zeta^4\right) 
+ k_3 \chi \left(\sigma^2 - \delta^2\right) \zeta \nonumber \\
& - k_4 \chi^4 
- \frac{1}{4} \chi^4 \ln\left(\frac{\chi^4}{\chi_0^4}\right) \nonumber \\
& + \frac{\xi}{3} \chi^4 \ln \left[ 
\left(\frac{(\sigma^2 - \delta^2)\zeta}{\sigma_0^2 \zeta_0}\right) 
\left(\frac{\chi^3}{\chi_0^3}\right) 
\right].
\end{align}
The above equation includes the last two terms to ensure the trace of the energy-momentum tensor is proportional to the fourth power of the dilaton field \(\chi\). The parameter
\(\xi\) is determined from the QCD \(\beta\)-function at one loop level for three colors and three flavours \cite{Papazoglou:1998vr}. The \(\pi\) meson mass, \(m_\pi\), \(K\) meson mass, \(m_K\), and the averaged mass of \(\eta\) and \(\eta'\) mesons are utilized to calculate the values of the parameters \(k_0, k_1, k_2, k_3\), and \(k_4\). The vacuum expectation values of scalar meson fields \(\sigma\) and \(\zeta\), denoted by \(\sigma_0\) and \(\zeta_0\), are expressed using the kaon decay constant \(f_K\) and pion decay constant \(f_\pi\) as \(\sigma_0=-f_\pi\) and \(\zeta_0=\frac{1}{2}(f_\pi-2f_K)\) \cite{Kumar:2023owb}. For vector mesons, the self-interaction term \(\mathcal{L}_{VV}\) is given by 
\begin{align}
\mathcal{L}_{VV} =\ & 
\frac{1}{2} \frac{\chi^2}{\chi_0^2} \left( m^2_\omega\, \omega^2 + m^2_\rho\, \rho^2 \right) \nonumber \\
& + g_4 \left( \omega^4 + 6\,\omega^2 \rho^2 + \rho^4 \right).
\end{align}
The vacuum value of \(\chi_0\) and the coupling constant \(g_4\) are fitted to the effective nucleon mass \cite{Wang:2001jw}. To account for the non-zero masses of pseudoscalar mesons, the explicit symmetry-breaking term \(\mathcal{L}_{\chi SB}\) is incorporated as \cite{Papazoglou:1998vr, Wang:2002aq}
\begin{eqnarray}
\mathcal{L}_{\chi SB} & = &-  \frac{\chi^2}{\chi_0^2} 
\left[ m_\pi^2 f_\pi \sigma + 
\left(\sqrt{2} m_K^2 f_K -\frac{ m_\pi^2}{\sqrt{2}} f_\pi \right)\zeta \right]
\qquad
\end{eqnarray}
The sixth term in Eq.~\eqref{eqn:2} is incorporated via an explicit symmetry-breaking term to achieve a physically accurate value for the strange quark mass \(m_s\), where $S_1$ is the strange quark matrix operator and is defined as 
$S_1=\frac{1}{3}\left(I-\lambda_8 \sqrt{3}\right)=diag(0,0,1)$ \cite{Wang:2001jw, Wang:2002aq}. Finally, the last term \(\mathcal{L}_{c}\) corresponds to the confinement potential term representing confinement of quarks within baryons and is written as 
\begin{eqnarray}
\mathcal{L}_c=-\bar\Psi \chi_c \Psi.
\end{eqnarray}
Dirac equation for a quark field $\Psi_{qi}$ , under the presence of meson mean fields becomes
\begin{eqnarray}
\left[- i \boldsymbol{\alpha} \cdot \nabla + \chi_c(r) + \beta m^*_q \right]\Psi_{qi} = e^*_q \, \Psi_{qi},
\end{eqnarray}
where the subscript $q$ denotes the quarks within a baryon of type $i$ (where $i$=$n$, $p$) and $\boldsymbol\alpha$, $\beta$ are the standard Dirac matrices. The effective quark mass $m_q^*$ and energy $e_q^*$ are expressed in terms of relevant coupling constants, and the scalar and vector meson fields as \cite{Kumar:2023owb}
\begin{eqnarray}
m^*_q & = & - g^q_\sigma \sigma - g^q_\zeta \zeta - g_\delta^q I^{3q} \delta + \Delta m,
\label{eqn:3}
\end{eqnarray}
and
\begin{eqnarray}
e^*_q & = & e_q - g^q_\omega \omega - g_\rho^q I^{3q} \rho.
\end{eqnarray}
In Eq.~\eqref{eqn:3}, to fit the vacuum masses of the quarks, the value of \(\Delta m\) is taken as \(77\) MeV for the \(s\) quark, whereas, for \(u/d\) quarks \(\Delta m = 0 MeV\). Here, \(I^{3q}\) refers to the third component of the isospin quantum number.
The effective mass of the $i^{th}$ baryon in terms of in-medium spurious center-of-mass momentum $\langle p^{*2}_{i\:\text{cm}} \rangle$ \cite{Barik:1985rm, Barik:2013lna} and the effective energy of the constituent quarks $e^*_q$ can be written as
\begin{eqnarray}
M^*_i & = & \sqrt{\left( \sum_q n_{qi} e^*_q + E_{i\:\text{spin}}\right)^2 - \langle p^{*2}_{i\:\text{cm}} \rangle}.
\end{eqnarray}
where, $n_{qi}$ represents the number of quarks of type $q$ within the $i^{th}$ baryon. Additionally, the in-medium spurious center-of-mass correction $\langle p^{*2}_{i\:\text{cm}} \rangle$  is given by
\begin{eqnarray}
\langle p^{*2}_{i\:\text{cm}}\rangle & = & \sum_q\frac{\left( 11 e^*_q + m^*_q \right)}{ 6\left( 3 e^*_q + m^*_q \right)} \left( e^{*2}_q - m^{*2}_q \right).
\end{eqnarray}

\subsection{Light front quark model}
\label{Sec2.2}
In LFQM \cite{Choi:1997iq, Choi:1999nu, Choi:2015ywa}, mesons are described as relativistic bound states consisting of a quark and an antiquark. The model employs the light-front framework, which is notably helpful for modeling hadronic systems while maintaining special relativity, especially when there are boosts along the direction of motion. The Fock state is described using a noninteracting \(q \bar q\) representation, with the interactions integrated into the mass operator \( M= H_0 + V_{q\bar{q}} \), to preserve compliance with the group structure satisfying the Poincaré algebraic commutation relations \cite{Keister:1991sb}.

\par The light-front wave function (LFWF), \( \boldsymbol{\Psi}_{q\bar{q}} \), encodes the full dynamics of the quark-antiquark system within the meson and is obtained by solving the eigenvalue equation of a QCD-inspired effective Hamiltonian, given by \cite{Choi:1999nu, Arifi:2024mff} 
\[
M |\boldsymbol{\Psi}_{q\bar{q}}\rangle = (H_0+ V_{q \bar{q}})|\boldsymbol{\Psi}_{q\bar{q}}\rangle =  M_{q\bar{q}} |\boldsymbol{\Psi}_{q\bar{q}}\rangle,
\]
where \(M_{q \bar{q}}\) is the mass eigenvalue of the meson. The relativistic kinetic energy \(H_0\) and the effective potential \(V_{q \bar{q}}\), which includes the linear confining potential \(V_{conf}\), the Coulomb potential of one-gluon exchange \(V_{coul}\) and the hyperfine potential \(V_{Hyp}\) are given as 
\begin{eqnarray}
H_0=\sqrt{m_q^{*2} + \textbf{p}_q^2} + \sqrt{m_{\bar{q}}^{*2} + \textbf{p}_{\bar{q}}^2},
\end{eqnarray}
and
\begin{eqnarray}
V_{q\bar {q}}=
\underbrace{a + b r}_{\text{conf}} 
- \underbrace{\frac{4 \alpha_s}{3r}}_{\text{coul}} 
+ \underbrace{\frac{32\pi \alpha_s \, \langle\mathbf{S}_q \cdot \mathbf{S}_{\bar{q}}\rangle}{9 m^*_q m^*_{\bar{q}}} \delta^{(3)}(r)}_{\text{hyp}},
\end{eqnarray}
respectively. In the above, \(a\) and \(b\) are the linear confining potential parameters, \(\alpha_s\) is the running coupling constant, and the spin interaction term \( \langle \mathbf{S}_q \cdot \mathbf{S}_{\bar{q}} \rangle \) takes the value \( \frac{1}{4} \) for vector mesons and \( -\frac{3}{4} \) for pseudoscalar mesons. Studies indicate that \(\alpha_s\) varies with density and temperature \cite{Arifi:2023jfe, Schneider:2003uz}; recent findings in Holographic QCD also investigate the changes in \(\alpha_s\) within the quark phase \cite{Arefeva:2025okg}. However, in this study, we focus on \(\rho\) mesons within nuclear matter and, for simplicity, consider \(\alpha_s\) to be a constant.
\par The meson state \( |M(\textbf{P}, \textit{J}, \textit{J}_z)\rangle \), expressed as a bound state of quark (\(q\)) and antiquark (\(\bar{q}\)) having total momentum \(\textbf{P}\) and total angular momentum \((\textit{J}, \textit{J}_z\)) can be written as
\begin{align}
| \mathcal{M}(\textbf{P}, \textit{J}, \textit{J}_z) \rangle 
&= \int \left[ \mathrm{d}^3 \mathbf{p}_q \right] 
         \left[ \mathrm{d}^3 \mathbf{p}_{\bar{q}} \right] \, 
         2(2\pi)^3 \delta^3(\mathbf{P} - \mathbf{p}_q - \mathbf{p}_{\bar{q}}) \notag \\
&\quad \times \sum_{\lambda_q, \lambda_{\bar{q}}} 
\Psi^{\textit{J} \textit{J}_z}_{\lambda_q \lambda_{\bar{q}}}(x, \mathbf{k}_\perp) \,
| q_{\lambda_q}(p_q) \, \bar{q}_{\lambda_{\bar{q}}}(p_{\bar{q}}) \rangle.
\end{align}
here \(p_j\) and \(\lambda_j\) are the momentum and helicity of quark (\(j=q\)) and antiquark (\(j=\bar{q}\)), with the LF momentum defined as \(\textbf{p}_j = (p_j^+, \, \mathbf{p}_{j\perp})\) and \(\mathrm{d}^3 \textbf{p}_j \equiv \mathrm{d}p_j^+ \, \mathrm{d}^2 \mathbf{p}_{j\perp}/2(2\pi)^3\). The Lorentz-invariant internal variables (\(x, \textbf{k}_\perp\)) are then expressed as \( x_j = \frac{p_j^+}{P^+} \) and \( \mathbf{k}_{\perp j} = \mathbf{p}_{\perp j} - x_j \mathbf{P}_\perp \). The trial wave function in momentum space is given by \cite{Arifi:2024mff, Arifi:2022pal}
\begin{eqnarray}
\mathbf \Psi_{\lambda_q \lambda_{\bar{q}}}^{\textit{J} \textit{J}_z}(x, \mathbf{k}_\perp) &=& \mathbf \Phi(x, \mathbf{k}_\perp) \mathcal R_{\lambda_q \lambda_{\bar{q}}}^{\textit{J} \textit{J}_z}(x, \mathbf{k}_\perp).
\end{eqnarray}
Here, we use the lowest-order harmonic oscillator wave function for the radial part of the trial wave function, which is expressed as 
\begin{eqnarray}
\mathbf \Phi^*_{1S}(x, \mathbf{k}_\perp) &=& \frac{4\pi^{3/4}} {\beta^{3/2}} \sqrt{\frac{\partial
{k_z}^*}{\partial x}} e^{-\frac{\mathbf{k}^2}{2\beta^2}},
\end{eqnarray}
where \(\beta\) is the variational parameter governing the wavefunction scale in our mass spectroscopic analysis, provided that \(\alpha_s\) is constant for all mesons. While \(\beta\) can theoretically be modified in a medium, it is associated with the short-range characteristics of the meson wave function and is therefore expected to be largely unaffected by medium effects \cite{Arifi:2024tix}. Consequently, it does not influence the modifications of hadron properties. The variable transformation $\{k_z, \mathbf{\textbf{k}_\perp}\} \rightarrow \{x, \mathbf{\textbf{k}_\perp}\}$ involves the Jacobian,
\begin{eqnarray}
\frac{\partial k_z^*}{\partial x} & = & \frac{M^*_0}{4x(1 - x)} \left[ 1 - \frac{(m_q^{*2} - m_{\bar{q}}^{*2})^2}{M_0^{*4}} \right],
\end{eqnarray}
where $k_z^* = \left(x - 1/2\right) M^*_0 + (m_{\bar{q}}^{*2} - m_q^{*2})/2M^*_0$. It should be noted that additional calculations will be performed using the normalized LFWF \cite{Arifi:2023jfe}. Asterik $(*)$ indicates the medium dependence of the quantities. $ \mathcal R_{\lambda_q \lambda_{\bar{q}}}^{\textit{J} \textit{J}_z}(x, \mathbf{k}_\perp)$ represents the interaction-independent spin-orbit wave function and its explicit form (derived through the Melosh transformation) is given as \cite{Arifi:2024mff}
\begin{eqnarray}
\mathcal{R}^{*00}_{\lambda_q \lambda_{\bar{q}}} &=& 
\frac{1}{\sqrt{2} \mathcal{R}^*_0} 
\begin{pmatrix}
-k^L & \mathcal{A^*} \\
-\mathcal{A^*} & -k^R
\end{pmatrix}
\end{eqnarray}
and
\begin{eqnarray}
\mathcal{R}^{*11}_{\lambda_q \lambda_{\bar{q}}} &=& 
\frac{1}{\mathcal{R}^*_0} 
\begin{pmatrix}
\mathcal{A^*} + \frac{\mathbf{k}_\perp^2}{\mathcal{M^*}} & k^R \frac{\mathcal{M}^*_1}{\mathcal{M^*}} \\
-k^R \frac{\mathcal{M}^*_2}{\mathcal{M^*}} &  -  \frac{\left( k^R \right)^2}{\mathcal{M^*}}
\end{pmatrix}, \\[10pt]
\mathcal{R}^{*10}_{\lambda_q \lambda_{\bar{q}}} &=& 
\frac{1}{\sqrt{2} \mathcal{R}^*_0} 
\begin{pmatrix}
k^L \frac{\mathcal{M}^*_3}{\mathcal{M}^*} & \mathcal{A^*} + \frac{2\mathbf{k}_\perp^2}{\mathcal{M^*}} \\
\mathcal{A}^* + \frac{2\mathbf{k}_\perp^2}{\mathcal{M}^*} & -k^R \frac{\mathcal{M}^*_3}{\mathcal{M}^*}
\end{pmatrix}, \\[10pt]
\mathcal{R}^{*1-1}_{\lambda_q \lambda_{\bar{q}}} &=& 
\frac{1}{\mathcal{R}^*_0} 
\begin{pmatrix}
- \frac{\left(k^L\right)^2}{\mathcal{M}^*}  & k^L \frac{\mathcal{M}^*_2}{\mathcal{M}^*} \\
-k^L \frac{\mathcal{M}^*_1}{\mathcal{M}^*} & \mathcal{A}^* + \frac{\mathbf{k}_\perp^2}{\mathcal{M}^*}
\end{pmatrix},
\end{eqnarray}
for pseudoscalar and vector mesons, respectively. Here, \(\mathcal{R}^*_0= \sqrt{\mathcal{A}^{*2} + \mathbf{k}_\perp^2}\), \(k^{R(L)}= k_x \pm ik_y, \mathcal{A^*}=(1-x)m^*_q+xm^*_{\bar q}, \mathcal{M}^*=M^*_0+m^*_q+m^*_{\bar q}, \mathcal{M}^*_1=xM^*_0+m^*_q, \mathcal{M}^*_2=(1-x)M^*_0+m^*_{\bar q}\), \(\mathcal{M}^*_3=\mathcal{M}^*_2-\mathcal{M}^*_1\) and the boost invariant meson mass squared, \(M_0^*\) is given by
\begin{eqnarray}
M_0^{*2} &=& \frac{\mathbf{k}_\perp^2 + {m_q^*}^2}{x} + \frac{\mathbf{k}_\perp^2 + {m_{\bar{q}}^*}^2}{1 - x}.
\end{eqnarray}
Furthermore, spin-orbit wave functions \( \mathcal{R}^{*J J_z}_{\lambda_q \, \lambda_{\bar{q}}} \) are inherently constructed to satisfy the unitarity condition, \(\left\langle \mathcal{R}^{*J J_z}_{\lambda_q \lambda_{\bar{q}}} \middle| \mathcal{R}^{*J J_z}_{\lambda_q \lambda_{\bar{q}}} \right\rangle = 1.\)
However, the same spin wave functions can be derived from Dirac spinor products as
\begin{equation} \label{spin}
\mathcal{R}^{*JJ_z}_{\lambda_q\lambda_{\bar q}}  = \frac{1}{\sqrt{2} M^*_0}\bar{u}_{\lambda_q}^{}(p_q) \left[-\slashed{\epsilon}_{J_z} + \frac{(p_q-p_{\bar q})\cdot{\epsilon}_{J_z}}{M^*_0 + m_q^*+m^*_{\bar q}}\right] v_{\lambda_{\bar q}}^{}(p_{\bar q}).
\end{equation}
Here, $\bar{u}_{\lambda_q}^{}(p_q)$ and $v_{\lambda_{\bar q}}^{}(p_{\bar q})$ are the Dirac spinor. $\epsilon_{J_z}$ is the polarization vector. We have also found that the vector spin alignment $\langle \mathcal{R}^{*10}_{\lambda_q\lambda_{\bar q}} \rangle$ of the $\rho$ meson is $1/3$, which indicates the spin states are equally distributed \cite{Fu:2023qht}.

\section{Mass of meson}
\label{Sec3}
 \par To calculate the mass of the meson, we apply the variational principle using the QCD-inspired quark-antiquark Hamiltonian \(H_{q\bar{q}}\). This involves computing its expectation value with the chosen trial wavefunction $\mathbf \Phi^*_{1S}(x, \mathbf{k}_\perp)$, which has variational parameter \(\beta\) dependence. By minimizing the expectation value with respect to \( \beta \), we can fix this parameter and obtain the corresponding mass eigenvalue of the meson, given by
\[
M^*_{q\bar{q}} = \langle \Psi_{q\bar{q}} | H_{q\bar{q}} | \Psi_{q\bar{q}} \rangle = \langle \Phi^*_{1S} | H_{q\bar{q}} | \Phi^*_{1S} \rangle.
\]
To address the issue of negative infinity, the delta function \( \delta^3(\mathbf{r}) \) is replaced with a Gaussian smearing function, \(\delta^3(\mathbf{r}) \rightarrow \left(\frac{\sigma'^{3}}{\pi^{3/2}}\right) e^{-\sigma'^2 r^2},\) which effectively weakens the singularity \cite{Choi:2015ywa}. Finally, the expression for the in-medium mass of the meson is obtained as \cite{Arifi:2023jfe, Choi:2009ai} 
\begin{eqnarray}
M^*_{q\bar{q}} & = & \frac{\beta}{\sqrt{\pi}} \sum_{i=q, \bar{q}} z_i e^{z_i/2} \textbf{\textit{K}}_1\left(\frac{z_i}{2}\right) + a + \frac{2b} {\beta \sqrt{\pi}} -\frac{ 8 \alpha_s \beta}{3 \sqrt{\pi}} \nonumber\\
& & + \frac{32 \alpha_s \beta^3\langle \mathbf{S}_q \cdot \mathbf{S}_{\bar{q}} \rangle}{9 \sqrt{\pi} m^*_q m^*_{\bar{q}}},
\end{eqnarray}

where ${z_i} = m_i^{*2}/{\beta^2}$ and $\textbf{\textit{K}}_1$ is the modified Bessel function of the second kind. The constituent quark masses in free space are obtained from the CQMF model, while the potential parameters are determined by fitting the mass spectra of the $\eta'$ and $\rho$ mesons. The resulting values are summarized in Table~\ref{tab:1}.
\begin{table}[h!]
\renewcommand{\arraystretch}{1.5} 
\centering
\begin{tabular}{|c|c|c|c|c|c|}
\hline\hline
$m_{u/d}$ & $\beta$ & $\sigma'$ \cite{Dhiman:2019ddr} & $\alpha_s$ & $a$ & $b$ \cite{Arifi:2023jfe} \\
\hline
0.256 & 0.325 & 0.451 & 0.212 & -0.694 & 0.18 \\
\hline\hline
\end{tabular}
\caption{Constituent quark masses (in GeV) along with potential parameters \(\sigma', \alpha_s, a\) (in GeV) and \(b\) (in GeV\(^2\)).}
\label{tab:1}
\end{table}
\par The computed mass of the ground state \(\rho\) meson mass using the above obtained model parameters is given in Table~\ref{tab:2}. Compared to previous studies based on LFQM \cite{Choi:2007yu, Arifi:2023jfe}, the relativistic quark model \cite{Ebert:2006hj}, and experimental data from Particle Data Group (PDG) \cite{ParticleDataGroup:2024cfk}, the result obtained shows a significant level of agreement. 
\begin{figure*}[t!]
\includegraphics[scale=0.4]{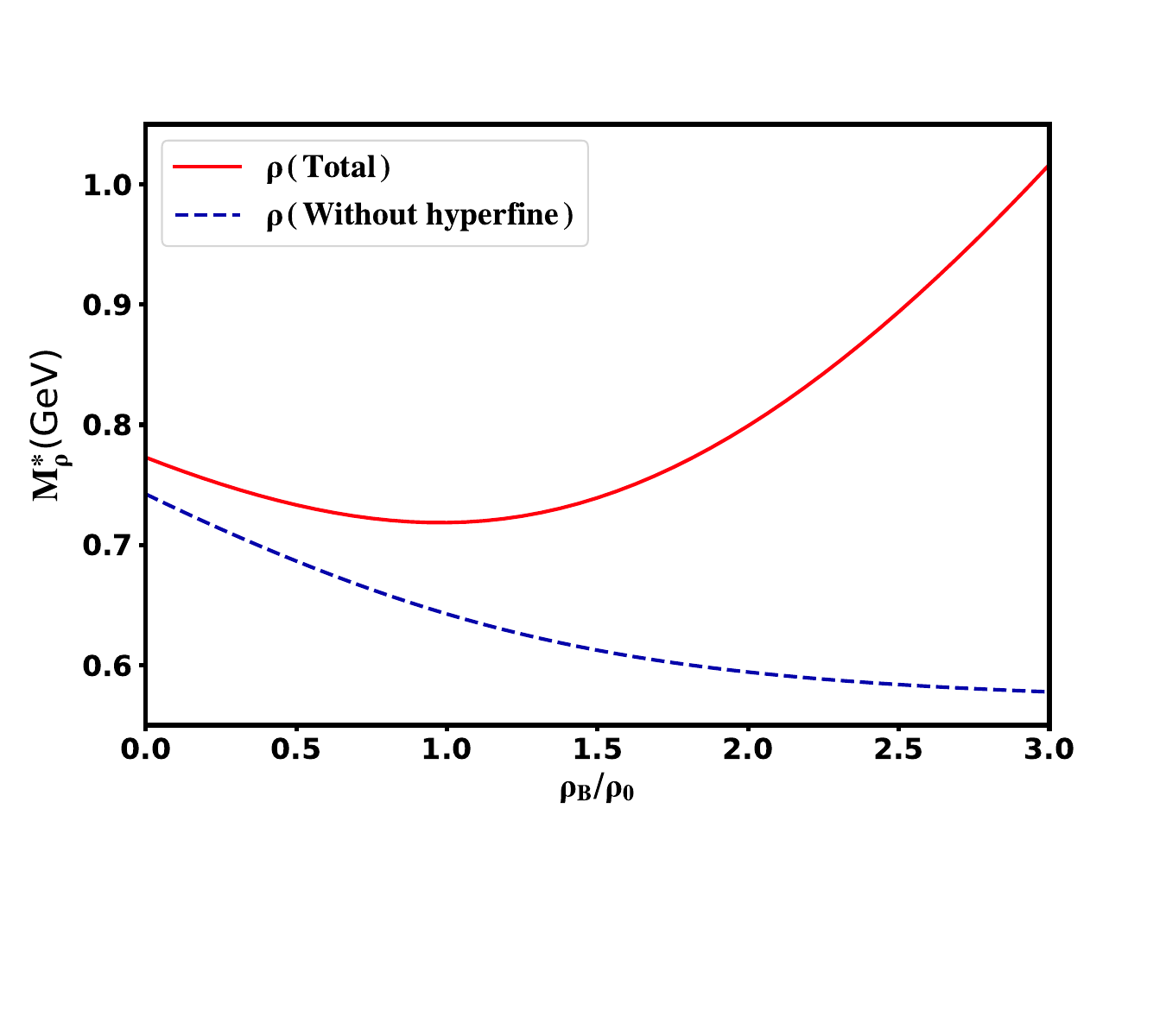}
\caption{Effective mass of \(\rho\) meson as a function of baryon density \(\rho_B\) (in units of nuclear saturation density \(\rho_0\)). Results are shown both with and without the contribution of the hyperfine potential.}\label{fig:1}
\end{figure*}
Figure~\ref{fig:1} illustrates the dependence of the \(\rho\) meson mass on the baryon density  $\rho_B$ (in units of nuclear saturation density $\rho_0$). The figure presents two curves corresponding to calculations performed with and without the inclusion of the hyperfine potential. The effective mass is observed to decrease at lower baryon densities up to $\rho_B=\rho_0$, followed by an increase after $\rho_B \ge \rho_0$. In contrast, the calculation that does not consider the hyperfine potential shows a consistent decline throughout the entire range of baryonic density. This comparison clearly indicates the significant role of the hyperfine potential in our calculations.
The unusual trend of mass decreasing and then increasing has also been observed in various model calculations \cite{Mishra:2014rha, Kumar:2018pqs, Mishra:2002uk}. A similar trend is evident in the CLAS data \cite{CLAS:2007dll}, which indicates that the mass of the \(\rho\) meson decreases for the \(C\) target and rises for the \(Fe-Ti\) target. At the baryon density of \(\rho_B=\rho_0\), the mass of the \(\rho\) meson is determined to be 0.719 GeV, compared to approximately 0.5 GeV as reported in Ref. \cite{deMelo:2018hfw}. A detailed comparison of the medium-modified masses at various baryon densities with different model predictions is presented in Table~\ref{tab:2}.

\begin{table*}[t]
\renewcommand{\arraystretch}{1.6} 
\centering
\begin{tabular}{|c|c|c|c|c|c|c|c|}
\hline\hline
\(\rho_B/\rho_0\)&  & \(M_\rho\) (GeV) &\(f_{\rho}^{\parallel *}\) (GeV) & \(f_{\rho}^{\perp *}\) (GeV) & \(f_{\rho}^{\parallel *}/f_{\rho}^{\perp *}\) &\(\Gamma^*\) (keV)\\
\hline
0&Our& 0.770& 0.219 &   0.176 & 1.242& 6.897 \\
&Exp.\cite{ParticleDataGroup:2024cfk}& 0.770 & 0.216& \dots&\dots&$7.04\pm0.06$\\
 &Ref.~\cite{Choi:2007yu} &0.770&0.215 & 0.173 &\dots& \dots\\
 &Ref.~\cite{Arifi:2023jfe}& 0.770 & 0.216 &\dots&\dots&\dots\\
 &Ref.~\cite{Ebert:2006hj} & 0.776 & 0.219 &\dots& \dots&\dots \\
 &Ref .~\cite{Mishra:2014rha}&0.753&\dots &\dots&\dots &\dots\\
  &Ref .~\cite{Shakin:1993ta}& 0.770&0.235&\dots&\dots &\dots\\
  &Ref .~\cite{Ball:2007zt}& \dots&0.216&0.165&1.309 &\dots\\
  &Ref .~\cite{Becirevic:2003pn}&\dots& \dots&\dots&1.428 &\dots\\
   &Ref .~\cite{Braun:2003jg}& \dots&\dots&\dots&1.348 &\dots\\
   &Ref .~\cite{Jansen:2009hr}&\dots& \dots&0.159&1.316 &\dots\\
    &Ref .~\cite{deMelo:2018hfw}&0.770&0.154 &\dots&\dots &\dots\\
    &Ref .~\cite{Hutauruk:2025bjd}&0.770&\dots &\dots&\dots &\dots\\
    &Ref .~\cite{Hatsuda:1991ez}&0.736&\dots &\dots&\dots &\dots\\
\hline
0.25& Our& 0.751  &0.219  &  0.173 & 1.266 & 7.116 \\
&Ref .~\cite{deMelo:2018hfw}&\dots&0.122 &\dots&\dots &\dots\\
&Ref .~\cite{Mishra:2014rha}&0.688&\dots &\dots&\dots &\dots\\
&Ref .~\cite{Hatsuda:1991ez}&0.712&\dots &\dots&\dots &\dots\\
\hline
0.5& Our& 0.733  & 0.219 & 0.169  & 1.293 & 7.278  \\
&Ref .~\cite{deMelo:2018hfw}& \dots& 0.109 &\dots&\dots &\dots\\
&Ref .~\cite{Mishra:2014rha}&0.635&\dots &\dots&\dots &\dots\\
&Ref .~\cite{Hutauruk:2025bjd}&0.726&\dots &\dots&\dots &\dots\\
&Ref .~\cite{Hatsuda:1991ez}&0.676&\dots &\dots&\dots &\dots\\
\hline
0.75&Our& 0.722  &0.218   &0.160   & 1.323 & 7.357\\
&Ref .~\cite{deMelo:2018hfw}&\dots&0.122 &\dots&\dots &\dots\\
&Ref .~\cite{Mishra:2014rha}&0.584&\dots &\dots&\dots &\dots\\
&Ref .~\cite{Hatsuda:1991ez}&0.639&\dots &\dots&\dots &\dots\\
\hline
1& Our& 0.719 & 0.217 & 0.160 & 1.358 & 7.342\\
  &Ref .~\cite{Mishra:2014rha}&0.0.538&\dots &\dots&\dots &\dots\\
  &Ref .~\cite{Shakin:1993ta}& 0.620 &0.282&\dots&\dots &\dots\\
  &Ref .~\cite{Hutauruk:2025bjd}&0.694&\dots &\dots&\dots &\dots\\
  &Ref .~\cite{Hatsuda:1991ez}&0.588&\dots &\dots&\dots &\dots\\
\hline
2&Our& 0.799 &0.213 & 0.142  & 1.496 & 6.307  \\
&Ref .~\cite{Mishra:2014rha}&0.382&\dots &\dots&\dots &\dots\\
&Ref .~\cite{Hutauruk:2025bjd}&0.658&\dots &\dots&\dots &\dots\\
&Ref .~\cite{Hatsuda:1991ez}&0.339&\dots &\dots&\dots &\dots\\
\hline
3&Our& 1.017 &0.208 &  0.130  & 1.600 & 4.761  \\
&Ref .~\cite{Mishra:2014rha}&0.276&\dots &\dots&\dots &\dots\\
\hline\hline
\end{tabular}
\caption{The in-medium weak decay constant (in units of GeV) for both longitudinal and transverse components, their ratio, and the decay width (in units of keV) of \(\rho\) meson have been compared with other model calculations up to baryonic density $\rho_B/\rho_0=3$.}
\label{tab:2}
\end{table*}

\begin{table*}[t]
    \renewcommand{\arraystretch}{1.7}

    \label{tab:3}
    \begin{tabular}{|l|l|cccc|l|cccc|l|}
        \hline\hline
        DAs & & $\langle \xi^2 \rangle$ & $\langle \xi^4 \rangle$ & $\langle \xi^6 \rangle$& $\langle \xi^8 \rangle$ & &$a_2$ & $a_4$ & $a_6$ & $a_8$ & $\langle x^{-1} \rangle$\\ \hline
        $\phi_{\rho}^{\parallel *} (x)$ 
            & Our &  0.195& 0.081 & 0.0403& 0.026 & Our & -0.015 & -0.018 & -0.009&-0.003 & 2.78 \\
             $(\rho_B/\rho_0=0)$&Ref.~\cite{Forshaw:2010py}  & 0.227 & 0.105 & 0.062 & 0.041 &Ref.~\cite{Ball:1998sk} & 0.18(10) & \dots & \dots & \dots& \dots\\
            & Ref.~\cite{Zhong:2023cyc} & 0.220 & 0.103 & 0.065& 0.046& Ref.~\cite{Zhong:2023cyc}&0.059 & \dots& \dots&\dots& \dots\\
            & 
            Ref.~\cite{Pimikov:2013usa} &  $0.216(21)$ & $0.089(9)$ &$0.048(5)$& 0.030(3)& Ref.~\cite{Pimikov:2013usa} &  0.047(58) & $-0.057(118)$ & \dots & \dots & 3.00\\
            & Ref.~\cite{Gao:2014bca} & 0.23 & 0.11 & 0.066 &  0.045 & Ref.~\cite{Gao:2014bca} & 0.092 & 0.031 & 0.015 & \dots& \dots\\
            & Ref.~\cite{Forshaw:2012im}&0.228&\dots&\dots&\dots&Ref.~\cite{Ball:2004rg} & $0.09
            $ & $0.03(2)$ & \dots & \dots& \dots\\
            & Ref.~\cite{Arifi:2025olq} & 0.193 &  0.078 & 0.040& \dots & Ref.~\cite{Ji:1992yf} & $-0.03$ & $-0.09$ &0.7 &\dots& \dots\\
            & Ref.~\cite{Arthur:2010xf} & 0.25(2)(2) & \dots & \dots & \dots& Ref.~\cite{Braun:2016wnx} & 0.132(27) & \dots & \dots & \dots& \dots\\
        $\rho_B/\rho_0=1$ 
            & Our & 0.209 & 0.092 & 0.052 &0.033 & Our & 0.027 & $-0.001$ & -0.005 &$-0.004$ & 3.02\\
        $\rho_B/\rho_0=2$ 
            & Our & 0.215 & 0.097 & 0.056 &0.037 & Our & 0.043 & $0.012$ & 0.003 &$0$ &3.22\\
        $\rho_B/\rho_0=3$ 
            & Our & 0.214 & 0.097 & 0.056 &0.037 & Our & 0.041 & $0.016$ & 0.007 &$0.003$ &3.30\\
            \hline 
        $\phi_{\rho}^{\perp *} (x)$ 
            & Our & 0.201 & 0.085 & 0.045 & 0.028 & Our & 0.004 & $-0.02$ & $-0.012$ & -0.004 & 2.84\\
           $(\rho_B / \rho_0 = 0)$ & Ref.~\cite{Arifi:2025olq} & 0.202 & 0.084 & 0.044 & $\dots$ & Ref.~\cite{Arifi:2025olq} & 0.007 & $-0.032$ & $-0.020$ & $\dots$ & $\dots$ \\

        $\rho_B/\rho_0=1$ 
            & Our & 0.219 & 0.099 & 0.057 & 0.037 & Our & 0.055 & 0.004 & $-0.005$ & $-0.005$ & 3.15\\
        $\rho_B/\rho_0=2$ 
            & Our & 0.226 & 0.105 & 0.062 &0.042 & Our & $0.076$ & $0.023$ & $0.007$ & 0.001 & 3.43\\
        $\rho_B/\rho_0=3$ 
            & Our & 0.226 & 0.106 & 0.063 &0.043 & Our & 0.075 & $0.029$ & $0.013$ & 0.006 &3.56\\ 
        \hline\hline
        
    \end{tabular}
    \caption{The Mellin moment $\xi=(2x-1)^n$ and Gegenbauer moments $a_n$ up to $n=8$ for vacuum and in-medium DAs of the $\rho$ meson up to baryonic density $\rho_B/\rho_0=3$ have been compared with other model calculations at 1 GeV scale. }
    \renewcommand{\arraystretch}{1}
\end{table*}


\section{Weak Decay Constant and Decay width}
\label{Sec4}
The decay constant is a quantity defined by the matrix elements that characterize the connection between the vacuum and a meson state with four-vector momenta \(P\). The decay constants \( f_{V}^{\parallel} \) and \( f_{V}^{\perp} \) relate to vector mesons with longitudinal and transverse polarizations, respectively, and are defined as \cite{Dhiman:2019ddr}
\begin{eqnarray}
\langle 0 | \bar{q} \gamma^\mu q | V(P, J_z) \rangle 
&=& f_{V}^{\parallel} M_V \, \epsilon^\mu(J_z), \\
\langle 0 | \bar{q} \sigma^{\mu\nu} q | V(P, J_z) \rangle 
&=& i f_{V}^{\perp} \left[ \epsilon^\mu(J_z) P^\nu - \epsilon^\nu(J_z) P^\mu \right] \nonumber \\
&& \hphantom{=} \phantom{i f_V^{\perp} \left[ \epsilon^\mu(J_z) P^\nu - \epsilon^\nu(J_z) P^\mu \right]}. 
\end{eqnarray}
\noindent
where \( \epsilon^\mu(J_z) \) is the polarization vector and \( M_V \) is the mass of the vector meson. The explicit form of the in-medium weak decay constants of the vector meson is given by 
\begin{eqnarray}
f_{V}^{\parallel *} & = & 2\sqrt{6} \int_0^1 dx \int \frac{d^2 \mathbf{k}_\perp}{2(2\pi)^3} \frac{\mathbf{\Phi}^*(x, \mathbf{k}_\perp) }{\mathcal{R}^*_0} \mathcal{O}_\parallel,\\
f_{V}^{\perp *} & = & 2\sqrt{6} \int_0^1 dx \int \frac{d^2 \mathbf{k}_\perp}{2(2\pi)^3} \frac{\mathbf{\Phi}^*(x, \mathbf{k}_\perp) }{\mathcal{R}^*_0}\mathcal{O}_{\perp}.
\end{eqnarray}
where \(\mathcal{O}_{\parallel} = \mathcal{A^*}+\frac{2\textbf{k}^2_\perp}{\mathcal{M}^*}\) and \(\mathcal{O}_{\perp} = \mathcal{A^*}+\frac{\textbf{k}^2_\perp}{\mathcal{M}^*}\). The longitudinal momentum fraction $x_{q/\bar q}$ of the quark or antiquark in free space is related to its in-medium counterpart $x_{q/\bar q}^*$ through the following relation
\begin{widetext}
\begin{eqnarray}
x_{q/\bar q}^* =
\begin{cases}
\dfrac{E_q^* + g_\omega^q \omega + g_\rho^q I^{3q} \rho + k_q^{*3}}
{E_q^* + E_{\bar{q}}^* + g_\rho^q(I^{3q}-I^{3\bar q})\rho+ P^{*3}}
= \dfrac{x_q + (g_\omega^q \omega + g_\rho^q I^{3q} \rho)/P^+}{1 + (I^{3q}-I^{3\bar q})\rho /P^+}, 
& \text{for quark } q, \\[12pt]
\dfrac{E_{\bar{q}}^* - g^{q}_\omega \omega - g^{q}_\rho I^{3\bar{q}} \rho + k_{\bar{q}}^{*3}}
{E_q^* + E_{\bar{q}}^* + g_\rho^q(I^{3q}-I^{3\bar q})\rho + P^{*3}}
= \dfrac{x_{q} - (g^{q}_\omega \omega + g^{q}_\rho I^{3\bar{q}} \rho)/P^+}{1 + (I^{3q}-I^{3\bar q})\rho /P^+}, 
& \text{for antiquark } \bar{q},
\end{cases}
\end{eqnarray}
\end{widetext}

\begin{figure*}[t!]
\includegraphics[scale=0.35]{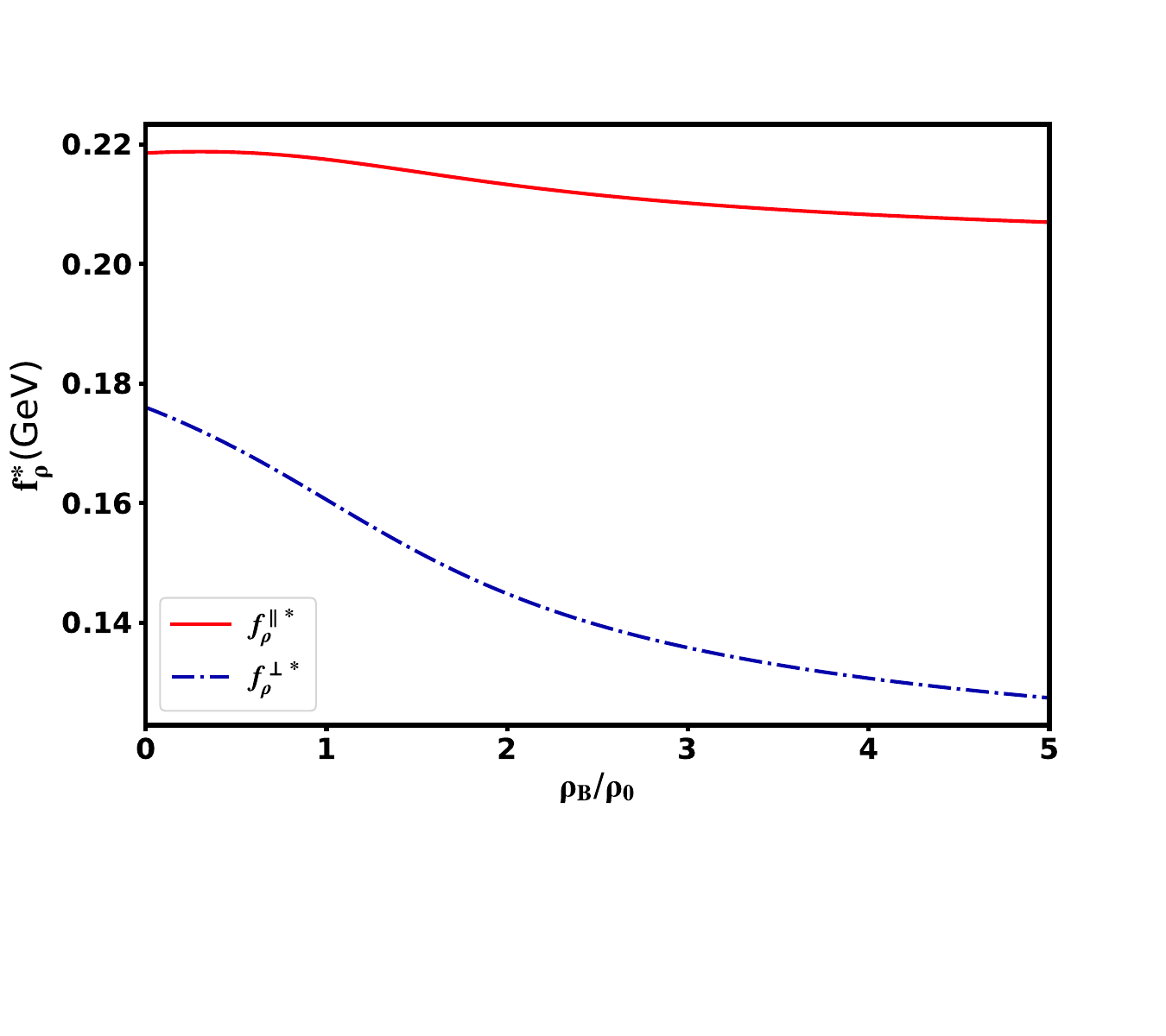}
\includegraphics[scale=0.35]{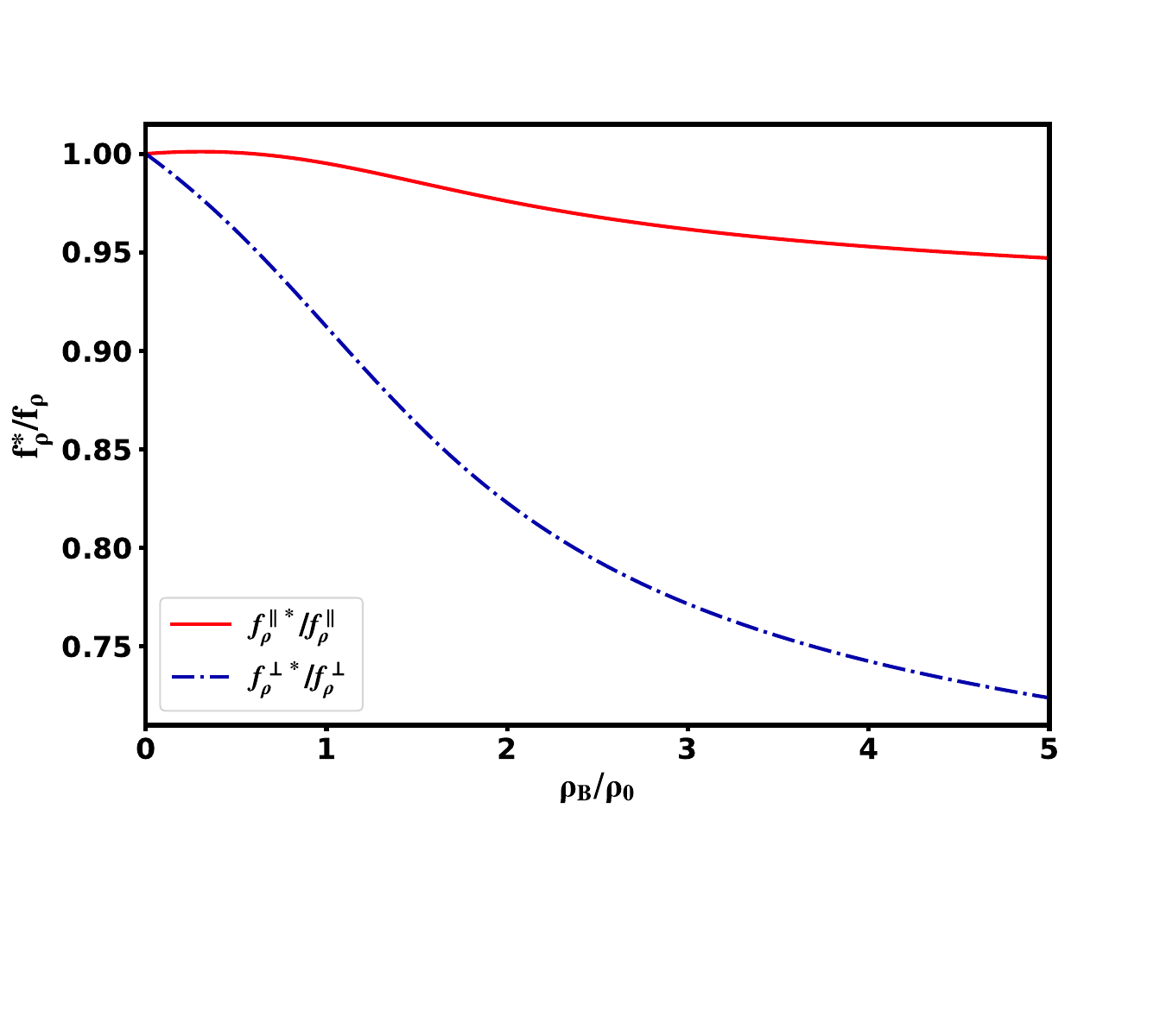}
\caption{(Color line)The in-medium $\rho$ meson decay constant (in units of GeV) has been plotted with baryonic density $\rho_B/\rho_0$ for both longitudinal and transverse components in the left panel. The ratio of in-medium and vacuum decay constant for both longitudinal and transverse polarization has been plotted with respect to baryonic density $\rho_B/\rho_0$ in the right panel.}\label{fig:2}
\end{figure*}

We have restricted our findings to \(x\) instead of \(x^*\) for the sake of simplicity. The vacuum longitudinal $f_{V}^{\parallel}$ and transverse $f_{V}^{\perp}$ decay constant is found to be $0.219$ GeV and $0.176$ GeV, which are been given in Table \ref{tab:2} along with comparison with PDG data \cite{ParticleDataGroup:2024cfk} and other theoretical model predictions \cite{Choi:2007yu, Arifi:2023jfe, Ebert:2006hj}. Our vacuum results are found to be in good agreement with other model predictions. The in-medium decay constant and normalized in-medium decay constant have been plotted with respect to baryonic density in the left and right panels of Fig. \ref{fig:2}, respectively. Both decay constants are found to be smoothly decreasing with increasing baryonic density. The longitudinal in-medium decay constant decreases with a slowly decreasing amplitude compared to the transverse decay constant, which can also be seen in the normalized in-medium decay constant case. Similar observation was also seen in Ref. \cite{deMelo:2018hfw}. The transverse decay constant $f_{V}^{\perp *}$ decreases by $25\%$ compared to the $4\%$ decrease for the longitudinal decay constant case at \(\rho_B=3\rho_0\), as shown in Table \ref{tab:2}. We have also observed that the ratio of the in-medium decay constant of longitudinal to transverse increases with increasing baryonic density. We have also compared our results for the in-medium decay constant with those in Refs. \cite{Shakin:1993ta,Ball:2007zt,Becirevic:2003pn,Braun:2003jg,Jansen:2009hr,deMelo:2018hfw} at vacuum as well as at finite baryon densities.
\par The decay process \( \rho^0 \rightarrow e^+ e^- \) is an important electromagnetic process and proceeds through an intermediate virtual photon. The corresponding medium-modified decay width is related to the longitudinal decay constant \( f_{V}^{\parallel *} \) and is given by \cite{deMelo:2018hfw}
\begin{equation}
\Gamma^*_{\rho^0 \rightarrow e^+ e^-} = \frac{4\pi \alpha^2 C_v^2}{3 M^*_\rho} (f_{\rho}^{\parallel*})^2,
\end{equation}
where \(C_v=1/\sqrt{2}\) for the \(\rho\) meson, \( \alpha \approx \frac{1}{137} \) is the fine-structure constant and \( M^*_\rho \) is the effective mass of the \(\rho\) meson.
\par
\begin{figure*}[t!]
\includegraphics[scale=0.35]{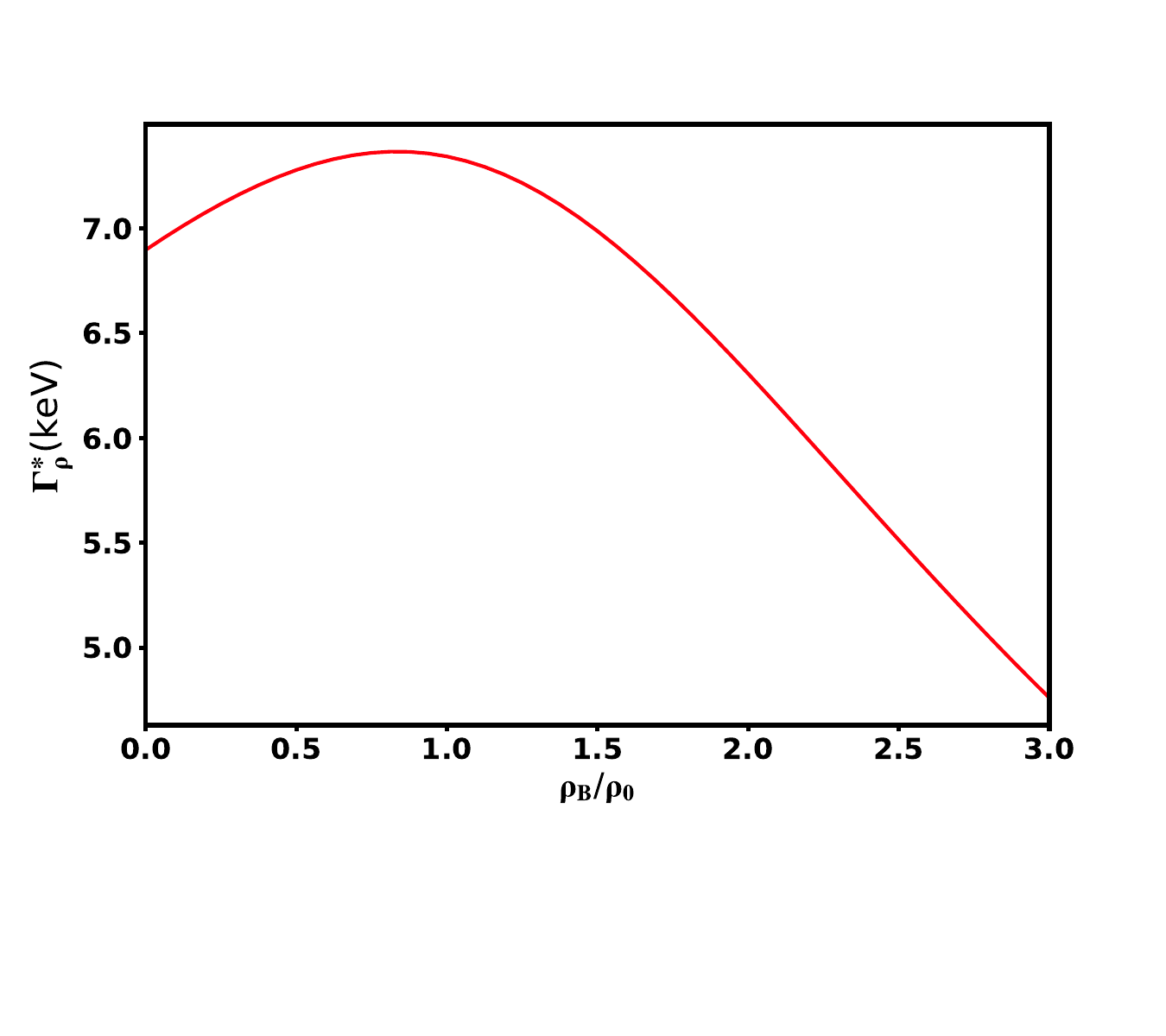}
\includegraphics[scale=0.35]{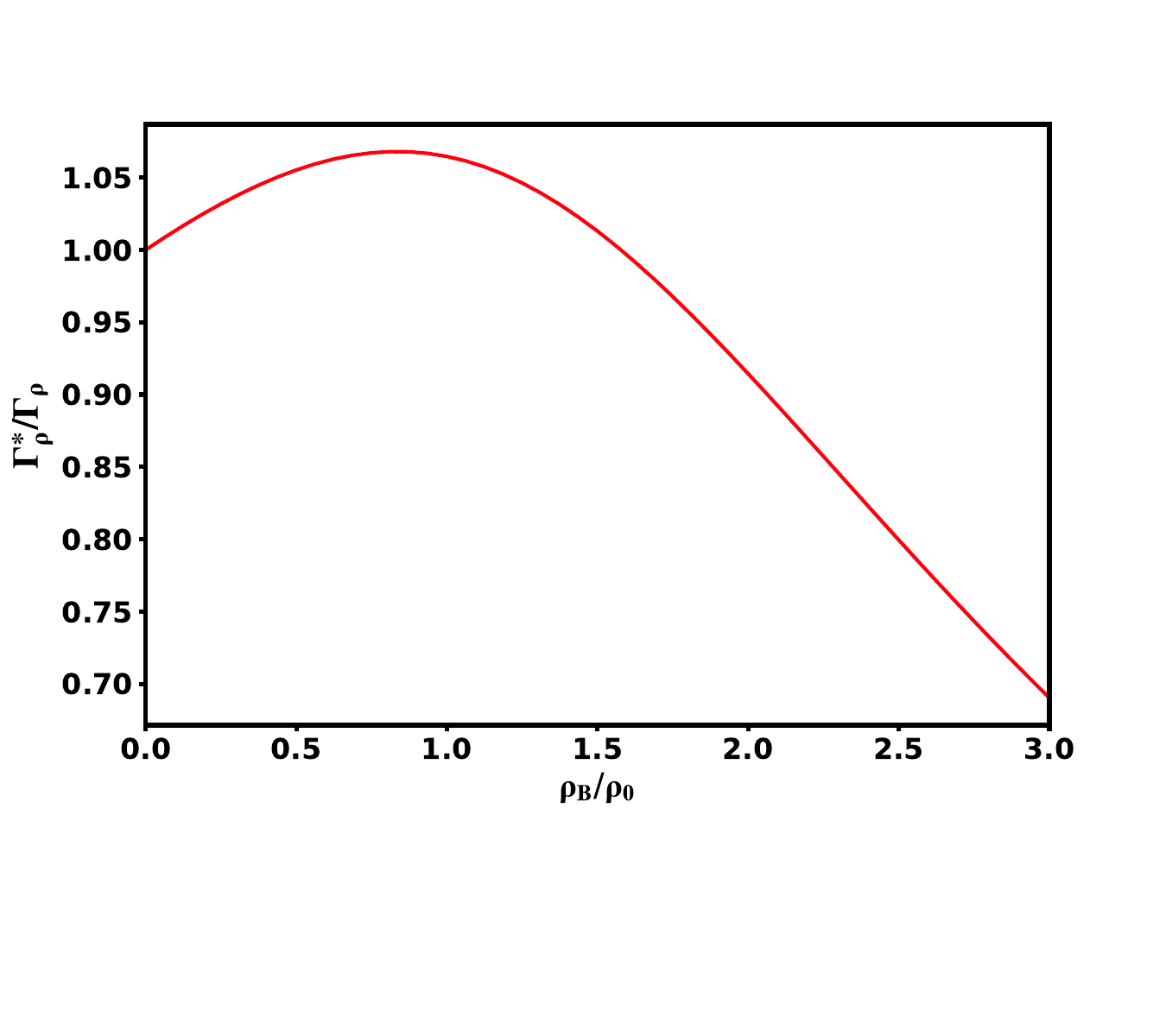}
\caption{(Color line) The in-medium and normalized decay width of \(\rho\) meson have been plotted with respect to baryonic density $\rho_B/\rho_0$ in units of keV in the left and right panel, respectively.}\label{fig:3}
\end{figure*}

 The vacuum decay width of $\rho^0 \to e^+e^-$ is found to be $6.897$ keV compared to the PDG data of $7.04\pm 0.06$ keV and has been presented in Table \ref{tab:2}. Our decay width is observed to deviate by about \(2\%\) from the PDG data \cite{ParticleDataGroup:2024cfk}. The in-medium decay width and its normalized form have been plotted with baryonic density in Fig. \ref{fig:3}. In contrast to the mass of the medium $\rho$ meson, the decay width demonstrates an opposing trend. The decay width increases up to $\rho_B=\rho_0$ and then starts to decrease at higher densities. However, in Ref. \cite{deMelo:2018hfw}, the decay width is shown to have only decreasing behavior. The observed decrease after $\rho_B=\rho_0$ is evident from the plot in Fig.~\ref{fig:3}, with specific values at different densities summarized in Table~\ref{tab:2}.
\par The branching fraction (\( \frac{\Gamma_i^*}{\Gamma_{\rm tot}}\)) for the decay of the $\rho^0 \to e^+e^-$ channel is determined by using the total decay width ($\Gamma_{\rm tot}=147.4$ MeV) of the $\rho^0$ meson \cite{ParticleDataGroup:2024cfk}. In vacuum, this leads to a branching fraction of \(4.7 \times 10^{-5}\) and associated lifetime (\(\tau = \frac{1}{\Gamma_{\rm tot}}\)), is found to be \(6.815\) x \(10^{-3}\) MeV\(^{-1}\). At baryonic density $\rho_B = \rho_0(3\rho_0)$, the partial width is adjusted for the medium, which results in an in-medium branching fraction of \(5.0 \times 10^{-5}(3.2 \times 10^{-5})\) and a lifetime of \(6.8 \times 10^{-3}(6.7 \times 10^{-3})\) MeV\(^{-1}\). These values indicate that both the branching fraction and the lifetime of the $\rho^0$ meson undergo only slight modifications in the nuclear medium, even at higher densities.


\section{Distribution Amplitude}
\label{Sec5}
DAs capture the essential non-perturbative dynamics of hadrons and play a central role in the study of exclusive processes in QCD \cite{Chernyak:1983ej, Lepage:1980fj}. These DAs are defined through matrix elements of nonlocal quark-antiquark operators, taken between the vacuum and the meson states \cite{Hwang:2010hw, Dhiman:2019ddr}
\begin{align}
\langle 0 | \bar{q}(0)\, \gamma^\mu\, q(0) | V(P, \lambda = 0) \rangle 
&= f_{V}^{\parallel} M_V\, \epsilon^\mu(\lambda)  \notag \\
&\quad \times \int_0^1 \phi_{V}^{\parallel}(x)\, dx, \\
\langle 0 | \bar{q}(0)\, \sigma^{\mu\nu}\, q(0) | V(P, \lambda = \pm1) \rangle 
&= i f_{V}^{\perp} \left[ \epsilon^\mu(\lambda) P_\nu - \epsilon^\nu(\lambda) P_\mu \right] \notag \\
&\quad \times \int_0^1 \phi_{V}^{\perp}(x)\, dx.
\end{align}
Here, \(\phi_{V}^{\parallel}\) and \(\phi_{V}^{\perp}\) are the twist-2 DAs of longitudinally and transversely polarized vector mesons in free space, respectively. The in-medium DAs for vector mesons are derived by integrating the LFWFs over the transverse momentum fraction as \cite{Dhiman:2019ddr}
\begin{eqnarray}
\phi_{V}^{\parallel *}(x) &=& \frac{2\sqrt{6}}{ f_{V}^{\parallel *}} \int \frac{d^2 \mathbf{k}_\perp}{2(2\pi)^3} \frac{\mathbf \Phi^*(x, \mathbf{k}_\perp)}{\mathcal{R}_0^*}\mathcal{O}_\parallel,\\
\phi_{V}^{\perp *}(x) &=& \frac{2\sqrt{6}}{ f_{V}^{\perp *}} \int \frac{d^2 \mathbf{k}_\perp}{2(2\pi)^3} \frac{\mathbf \Phi^*(x, \mathbf{k}_\perp)}{\mathcal{R}_0^*}\mathcal{O}_\perp.
\end{eqnarray}
\begin{figure*}[t!]
\includegraphics[scale=0.35]{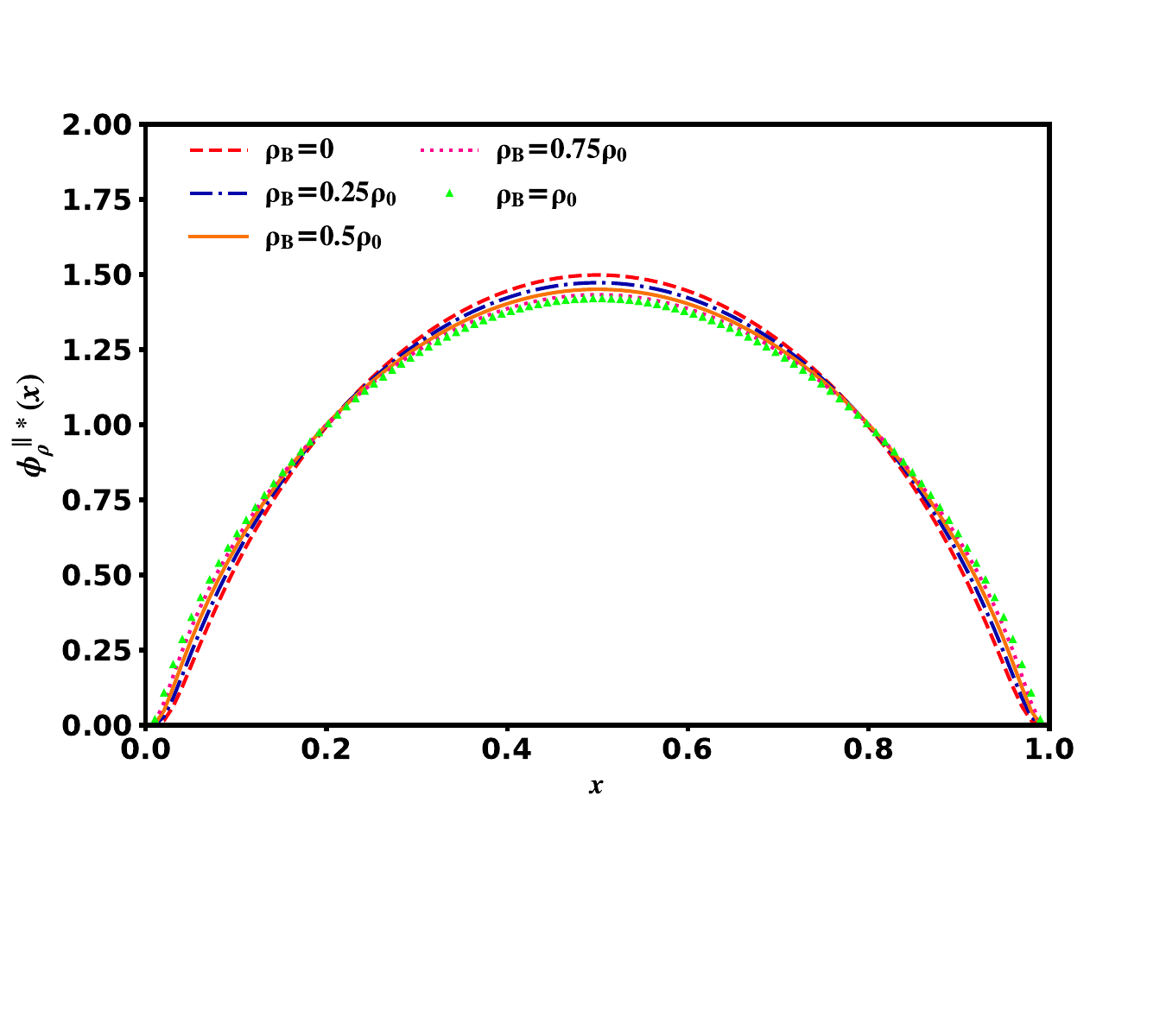}
\includegraphics[scale=0.35]{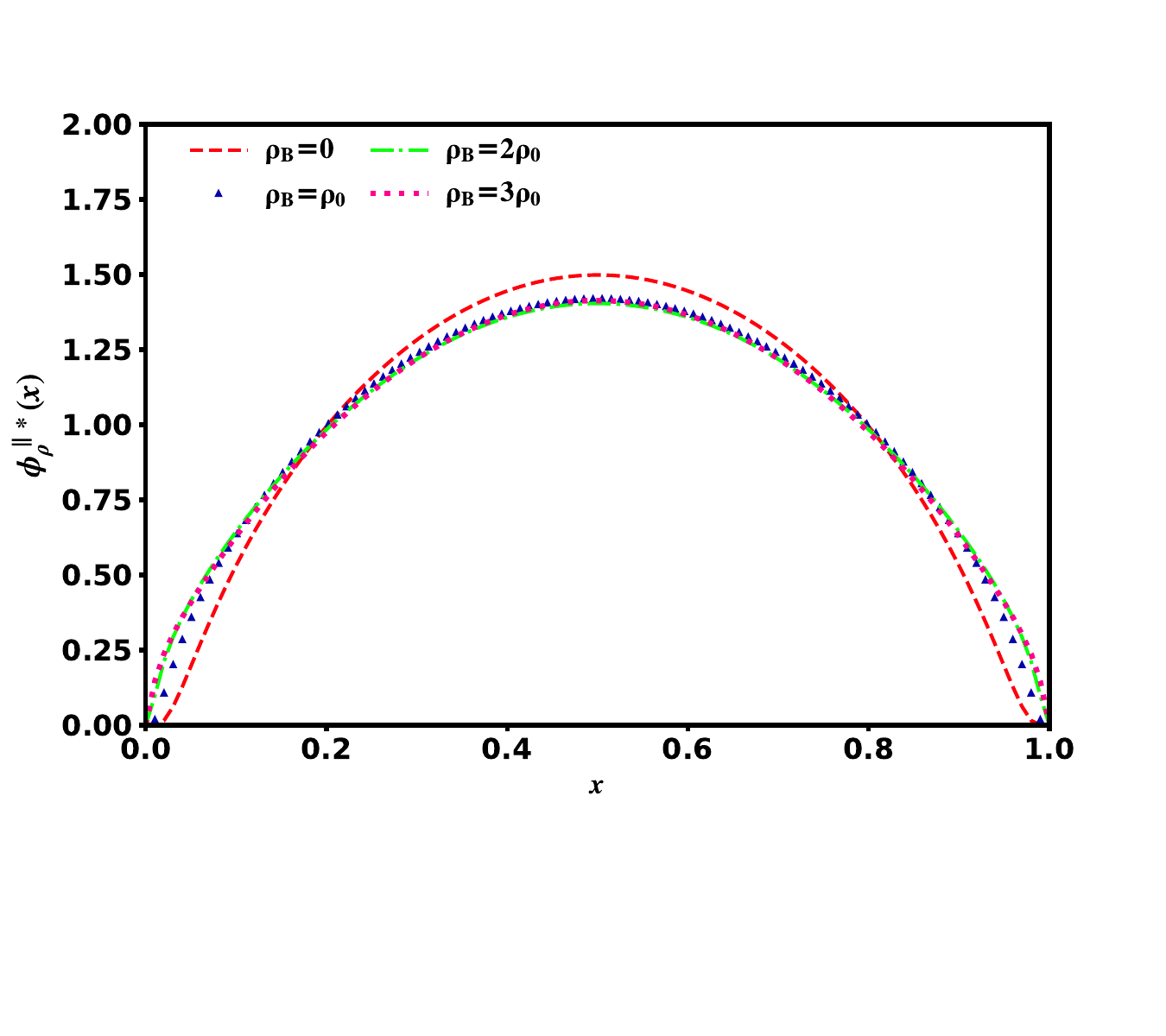}
\caption{(Color line) The in-medium longitudinal distribution amplitude has been plotted with respect to baryonic density $\rho_B/\rho_0$ in the interval of 0.25 in the left panel and in the interval of 1 in the right panel, respectively.}\label{fig:4}
\end{figure*}
\begin{figure*}[t!]
\includegraphics[scale=0.35]{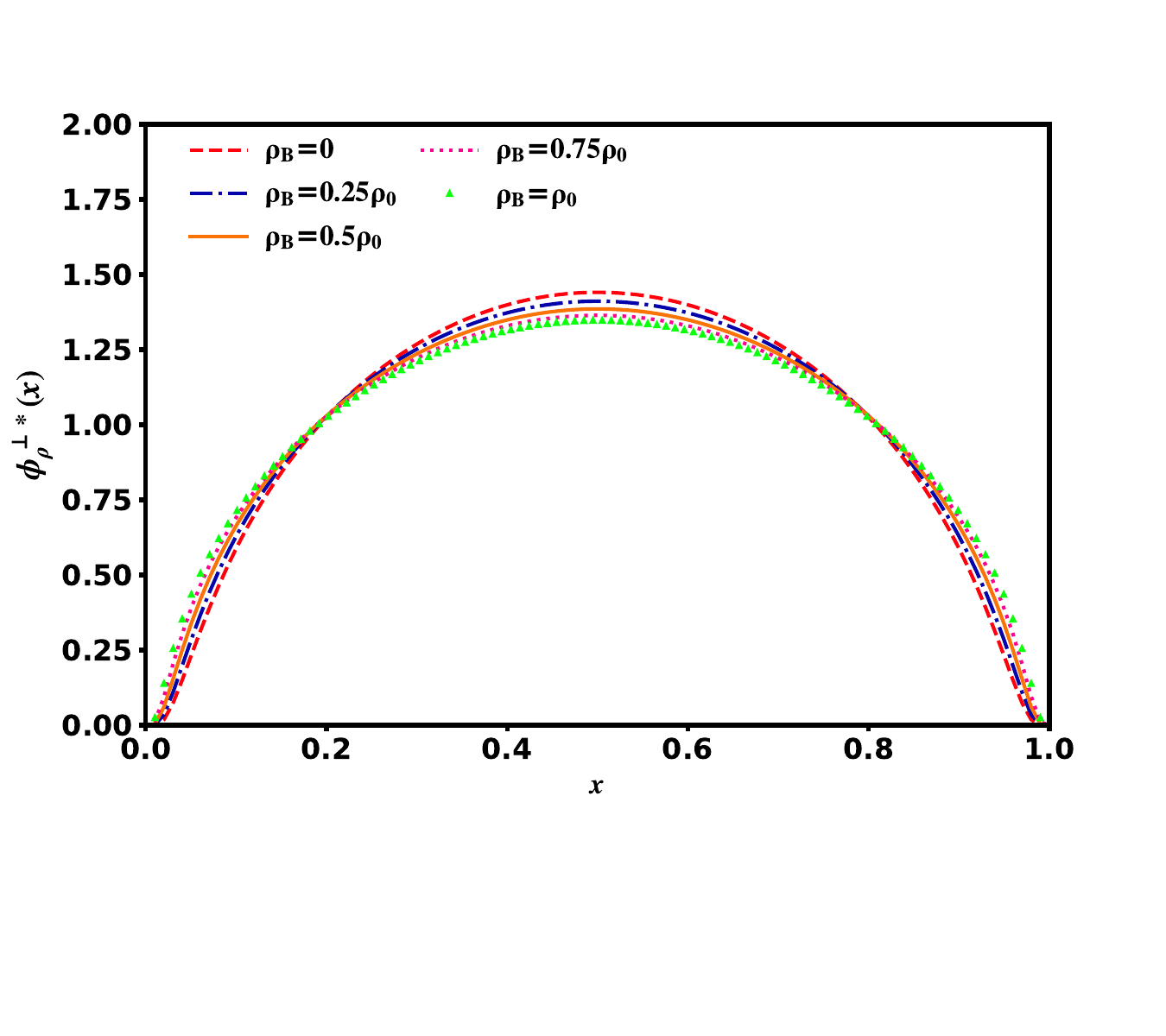}
\includegraphics[scale=0.35]{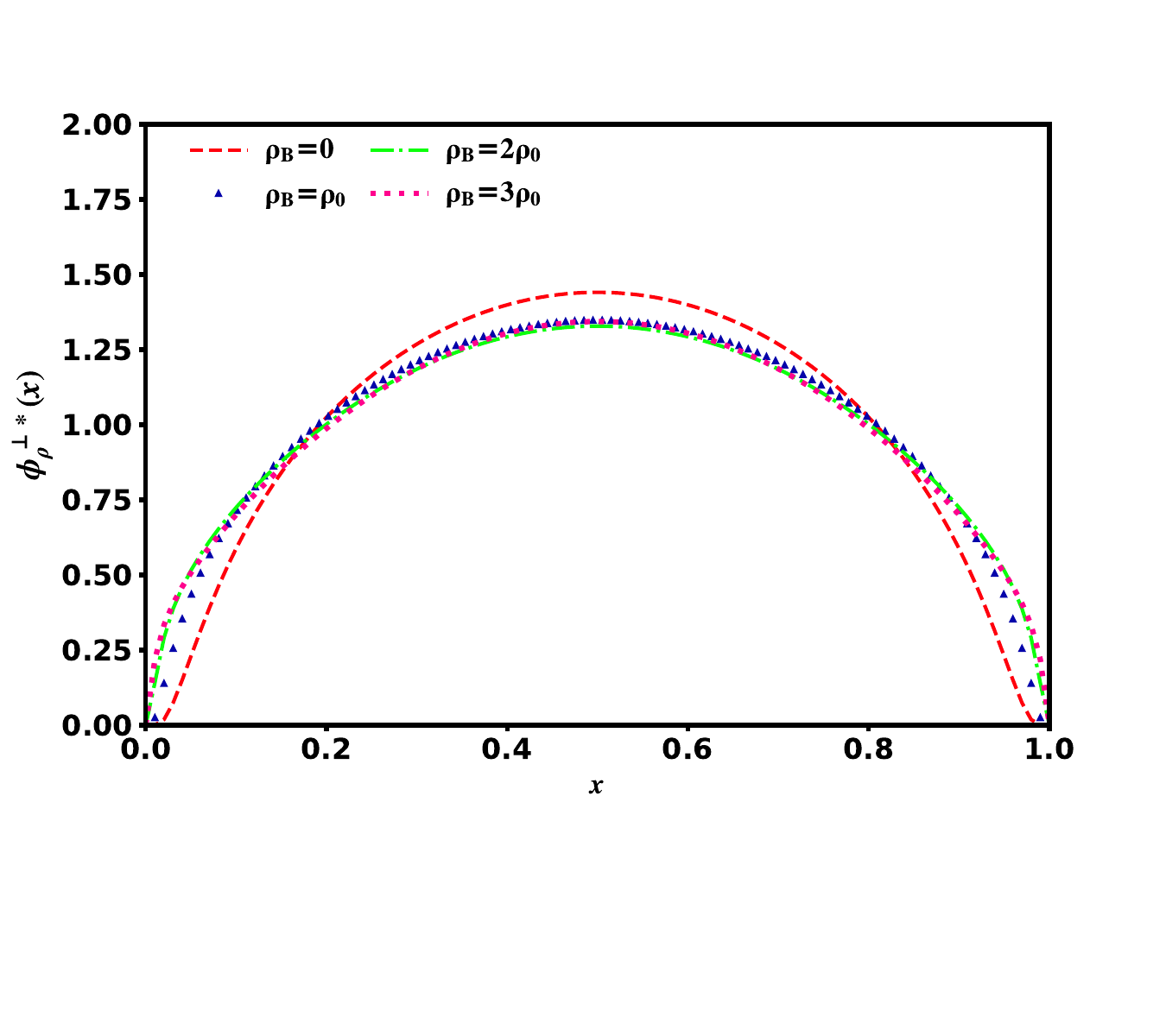}
\caption{The in-medium transverse distribution amplitude has been plotted with respect to baryonic density $\rho_B/\rho_0$ in the interval of 0.25 in the left panel and in the interval of 1 in the right panel, respectively}\label{fig:5}
\end{figure*}
\par Both the longitudinal and transverse DAs obeys the sum rule
\begin{eqnarray}
   \int dx \phi_{V}^{\perp *}(x)=\int dx\phi_{V}^{\parallel *}(x)=1,
\end{eqnarray}
at every scale. The longitudinal $\phi_{V}^{\parallel *}$ and transverse $\phi_{V}^{\perp *}$ DAs have been plotted at different baryonic densities in Fig. \ref{fig:4} and \ref{fig:5}, respectively, up to baryonic densities $3 \rho_0$. We observe that as the baryonic density increases, the distribution increases in the range $x \leq 0.225$ and $x \geq 0.775$, whereas it decreases in the intermediate range $0.225 \leq x \leq 0.775$ for both DAs. Similar kinds of results have also been obtained for the case of pion \cite{Puhan:2024xdq} and heavy pseudo-scalar mesons \cite{Arifi:2024mff}.
\par
\begin{figure*}[t!]
\includegraphics[scale=0.35]{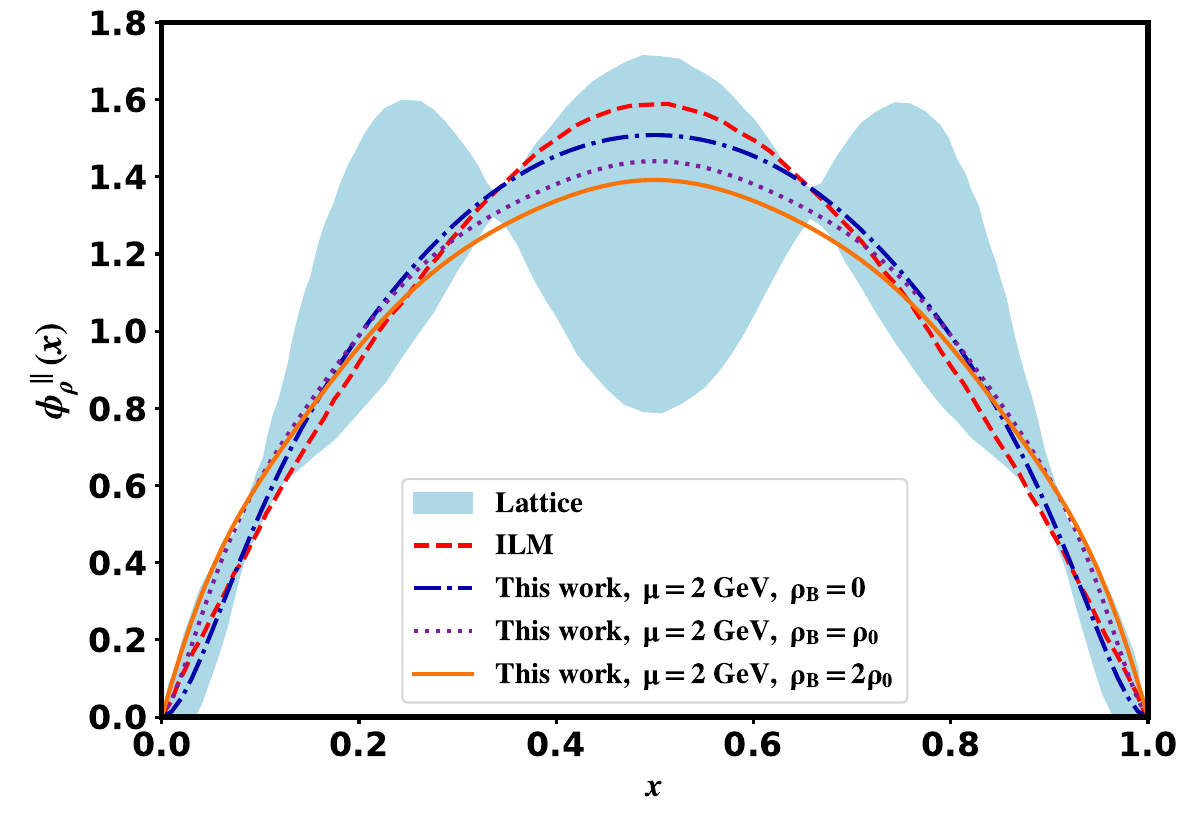}
\caption{The in-medium longitudinal distribution amplitude, which is evolved to \(\mu=2\) GeV, has been plotted with respect to baryonic density $\rho_B/\rho_0$ in the interval of 1. The result is compared with ILM and Lattice data \cite{Pimikov:2013usa,Liu:2023fpj}.}\label{fig:6}
\end{figure*}

 To have a comprehensive idea of in-medium DAs on a larger scale, we have evolved the DAs. This leading order (LO) QCD evolution of the DAs is carried out using the Efremov--Radyushkin-Brodsky-Lepage (ERBL) evolution equations \cite{Efremov:1979qk, Lepage:1980fj}. When expressed in the Gegenbauer polynomial basis, the evolution equations take on the form \cite{RuizArriola:2002bp} 
\begin{align}
\phi_{\rho}(x, \mu) &= 6x(1 - x) \sum_{n=0}^{\infty} C_n^{3/2}(2x - 1)\, a_n(\mu), \\
a_n(\mu) &= \frac{2(2n + 3)}{3(n + 1)(n + 2)} 
\left( \frac{\alpha_s(\mu)}{\alpha_s(\mu_0)} \right)^{\frac{\gamma_n^{(0)}}{2\beta_0}} \notag \\
&\quad \times \int_0^1 dx\, C_n^{3/2}(2x - 1)\, \phi_{\rho}(x, \mu_0), 
\end{align}

where \( C_n^{3/2}(2x - 1) \) is a Gegenbauer polynomial and \( \frac{\gamma_n^{(0)}}{2\beta_0} \) describes the anomalous dimension with
\begin{equation}
\gamma_n^{(0)} = -2 C_F \left( 3 + \frac{2}{(n+1)(n+2)} - 4 \sum_{m=1}^{n+1} \frac{1}{m} \right),
\end{equation}
and
\begin{equation}
\beta_0 = \frac{11}{3} c_A - \frac{2}{3} n_F,
\end{equation}
where \( c_A = 3 \) and \( n_F \) represent the number of active flavors. The \( c_F = \frac{4}{3} \) denotes the color factor, and \( \Lambda_{\text{QCD}} = 0.226~\text{GeV} \). Additionally, the strong coupling constant \( \alpha_s(\mu) \) is expressed by

\begin{equation}
\alpha_s(\mu) = \frac{4\pi}{\beta_0 \ln\left(\mu^2 / \Lambda^2_{\text{QCD}}\right)}.
\end{equation}
It is worth noting that the in-medium coupling constant has not been considered in this work. The longitudinal in-medium $\phi^{\rho}_{\parallel *}(x)$ has been compared with the lattice simulation results \cite{Pimikov:2013usa} and the Instanton Liquid Model (ILM) results \cite{Liu:2023fpj} in Fig. \ref{fig:6} at $\mu=2$ GeV. The initial model scale has been taken to be $\mu_0=1$ GeV for our calculations. The vacuum and in-medium evolved DAs are found to match with both of them. 

We can compute the \(n\)-th moment using the numerical integration of DAs over the light-front momentum fraction as \cite{Dhiman:2019ddr}
\begin{eqnarray}
\langle z_n \rangle &=& \int_0^1 dx\, z^n \phi^*(x),
\end{eqnarray}
where \(z\) is either \(\xi=(2x-1)\) or \(x^{-1}\). $z=x^{-1}$ is also called the inverse moment of the mesons. The values for all moments $\langle z_n \rangle$ and $a_n$ are summarized in Table \ref{tab:3}, together with comparisons to lattice simulations, PDG data, and various theoretical predictions at $\mu=2$ GeV. Our results in vacuum are observed to be comparable to other theoretical predictions. We found that the Mellin moments increase as the baryonic density increases.  In the same manner, the inverse moment also shows an increase with increasing baryonic density. The vacuum inverse moment is found to be $2.78$ and $2.84$ for longitudinal and transverse DAs, respectively. The inverse moment is nearly identical to that of the pion, which we reported previously in the light-cone quark model \cite{Puhan:2023ekt}. At a baryonic density of $3\rho_0$, the inverse moment is observed to increase by $15.75\%$ for both DAs. Overall, we established that baryonic density influences the DAs as well as their corresponding moments. 
\section{Form Factors}
\label{Sec6}
At the leading twist spin-1 vector meson case, there exist three electromagnetic form factors $(F_i(Q^2) (i=1,2,3))$, compared to one and two for the spin-0 and spin $1/2$ nucleon case. These Lorentz invariant form factors can be calculated by the matrix elements of $J^\mu$ current positioned between the initial $|M(P^+,\textbf{P}_\perp, S_z \rangle$ and final $\langle M(P^{+\prime},\textbf{P}^{\prime}_\perp, S_z'|$ state meson as follows \cite{Kumar:2019eck, Choi:2004ww}
\begin{align}
\langle M(P^{+\prime},\textbf{P}^{\prime}_\perp, S_z') | J^\mu | M(P^+, \mathbf{P}_\perp, S_z) \rangle= \quad\ \nonumber \\
\quad   - \epsilon^{\prime *} \cdot \epsilon\, (P + P')^\mu\, F_1(Q^2) \nonumber \\
  \quad + \left( \epsilon^{\prime *} \cdot n\, \epsilon \cdot (P + P') 
  - \epsilon \cdot n\, \epsilon^{\prime *} \cdot (P + P') \right) F_2(Q^2) \nonumber \\
 \quad + \frac{(\epsilon^{\prime *} \cdot n)(\epsilon \cdot n)}{2 M_v^2}\, (P + P')^\mu\, F_3(Q^2).
\label{electromagnetic}
\end{align}
Here, for the meson with spin projections $S_z$ and $S_z^{\prime}$, the four-momenta of the initial and final states are $P$ and $P'$, respectively. \(P_a=(P'+P)/2\) is the meson's average momentum. The momentum transferred between the initial and final meson states is represented by \(\Delta(\Delta^+,\Delta^-,\Delta_\perp)=P'-P\). Also, $Q^2=\Delta^2=-\Delta^2_\perp$. The medium modified charge, magnetic, and quadrupole form factors of vector mesons can be calculated using these form factors as
\begin{align}
G^{\ast}_C(Q^2) &= \left(1 + \frac{2}{3}\kappa \right) F^{\ast}_1(Q^2) + \frac{2}{3}\kappa\, F^{\ast}_2(Q^2) \nonumber\\ 
         &+ \frac{2}{3}\kappa(1 + \kappa)\, F^{\ast}_3(Q^2), \label{eq:GC} \\
G^{\ast}_M(Q^2) &= -F^{\ast}_2(Q^2), \label{eq:GM} \\
G^{\ast}_Q(Q^2) &= F^{\ast}_{1}(Q^2) + F^{\ast}_{2}(Q^2) + (1 + \kappa)\, F^{\ast}_{3}(Q^2). \label{eq:GQ}
\end{align}
Here, $\kappa=\frac{\Delta_\perp^2}{4 M^{\ast 2}_{q \bar q}}$. In this work, we have not taken into account the zero mode contributions as discussed in Ref. \cite{Kumar:2019eck, Choi:2004ww}. Additionally, we have noted that our model closely adheres to the asymptotic limit of QCD given in Refs. \cite{Brodsky:1992px, Hernandez-Pinto:2024kwg}, which states
\begin{eqnarray}
    G^{\ast}_C(Q^2):G^{\ast}_M
    (Q^2):G^{\ast}_Q(Q^2)  \;{\stackrel{\scriptstyle Q^2 \to \infty}{=}}\; 1-\frac{2 \kappa}{3}:2:-1.
\end{eqnarray}
From these in-medium form factors, one can determine the modified charge, magnetic moment ($\mu_{\rho}^*$), and quadrupole moment $Q_{\rho}^*$ of the \(\rho\) mesons in the symmetric nuclear medium at zero momentum transfer $Q^2=0$ as
\begin{eqnarray}
    eG^{\ast}_C(Q^2=0)=e, \ \ G^{\ast}_M(Q^2=0)&=\mu^{\ast}_{\rho}, \ \  \\ \nonumber
     G^{\ast}_Q(Q^2=0)=Q^{\ast}_{\rho}.
\end{eqnarray}
Here, $e$ denotes the static charge. $\mu^{\ast}_{\rho}$ and $Q^{\ast}_{\rho}$ are in Bohr magneton units $e/2 M^{\ast}_{\rho}$, and $e/M^{\ast2}_{\rho}$. $M_{\rho }$ refers to the physical mass of $\rho$ meson. The medium modified charge radii $\sqrt{\langle r_c^{\ast2}\rangle}$ of the corresponding vector mesons can be determined using the fundamental radius equation, given by
\begin{eqnarray}
    \langle r_c^{\ast2}\rangle &= & \frac{-6}{G^{\ast}_C(0)} \frac{\partial G^{\ast}_C(Q^2)}{\partial Q^2}\Big|_{Q^2\rightarrow0}.
\end{eqnarray}
Also, the $G^{\ast}_C(0)=1$.

\begin{figure*}[t!]
(a)\includegraphics[scale=0.3]{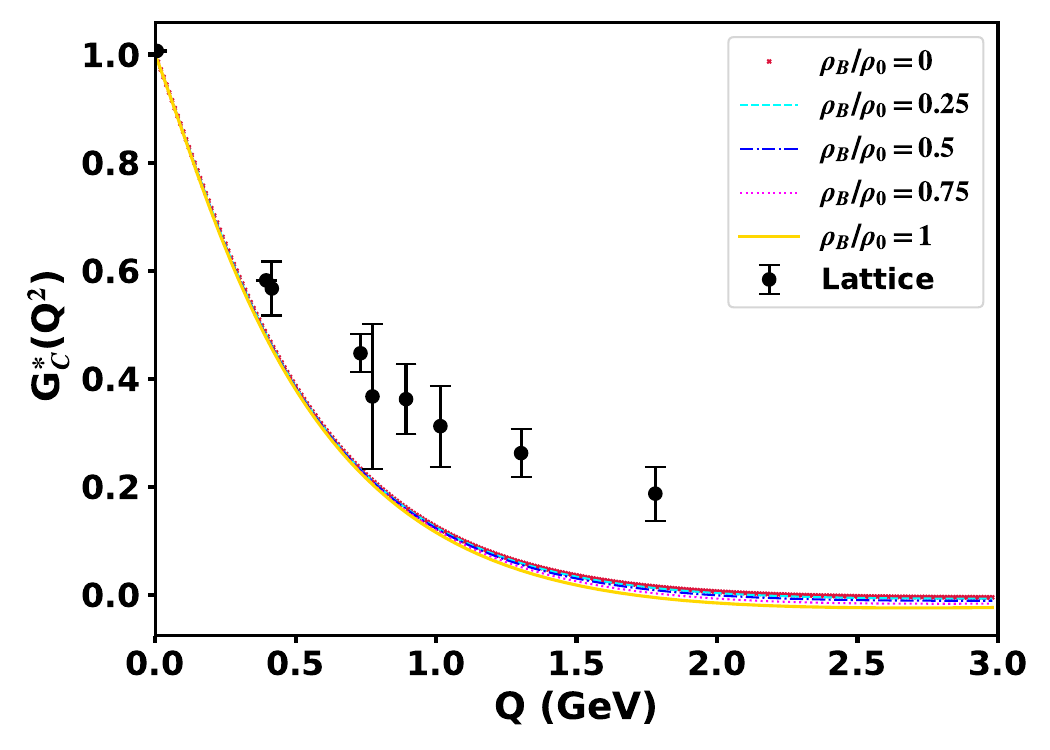}
(b)\includegraphics[scale=0.3]{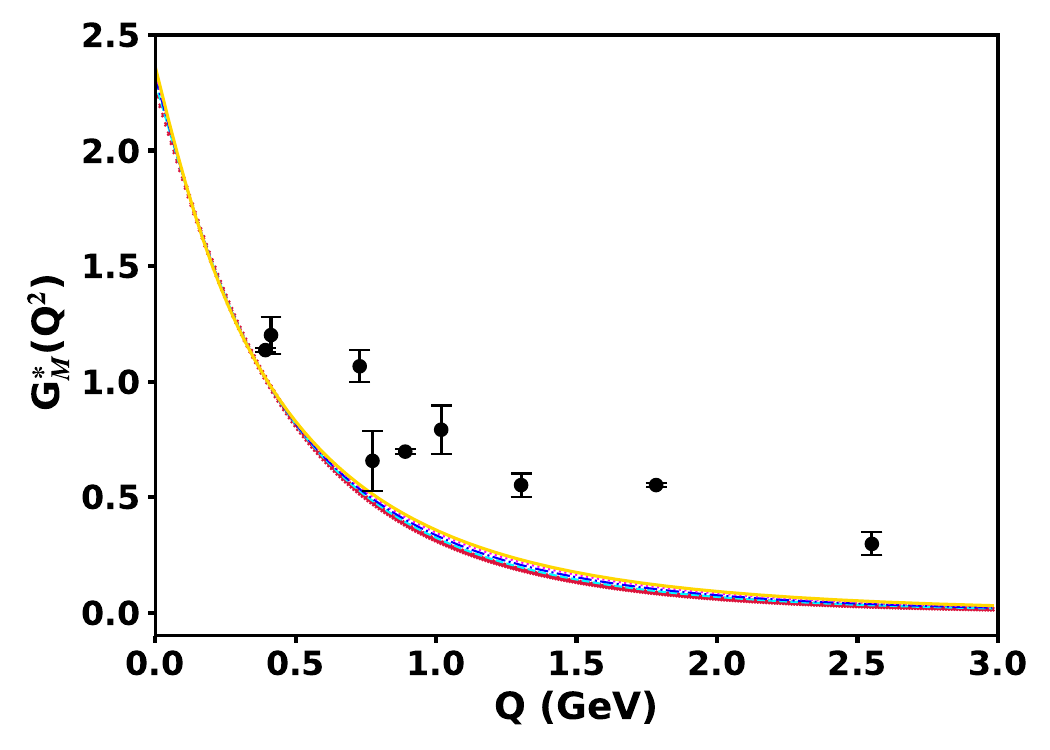}
(c)\includegraphics[scale=0.3]{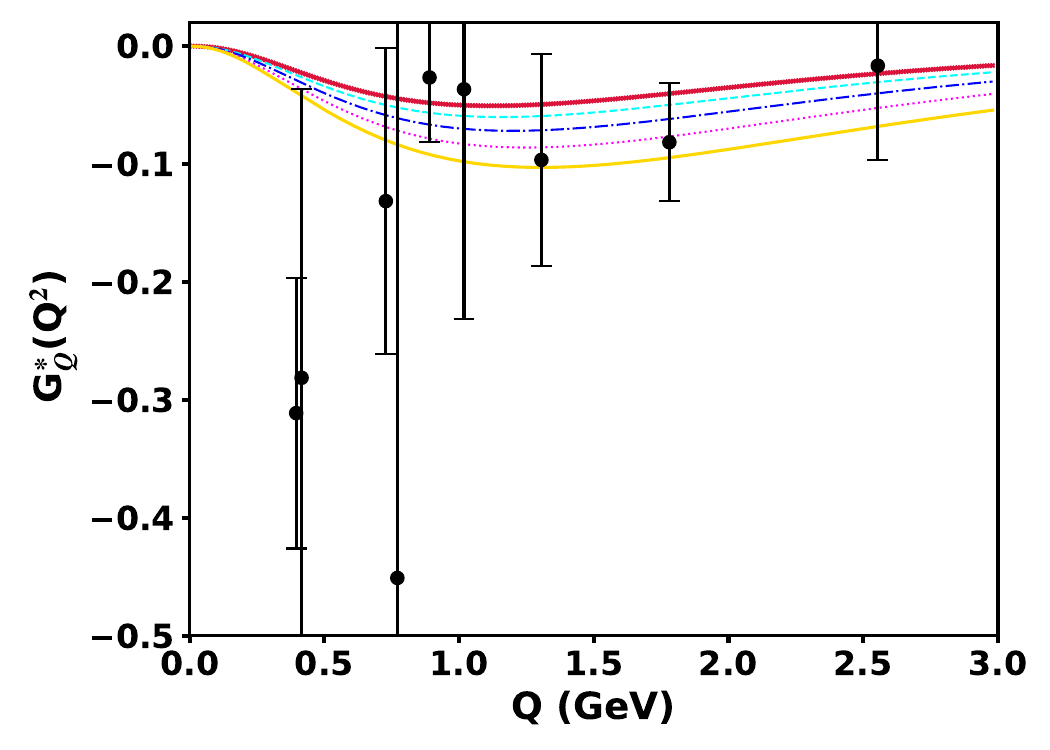}
\caption{ Variation of the medium modified (a) charge \((G^{\ast}_{C}(Q^2))\), (b) magnetic \((G^{\ast}_{M}(Q^2))\), and (c) quadrupole \((G^{\ast}_{Q}(Q^2))\) form factors with momentum transfer Q (GeV) shown at baryonic density ratios ($\rho_B/\rho_0$) varying from 0 to 1 at an interval of 0.25. Lattice simulation data are included for comparison \cite{Lasscock:2006nh}}\label{plot1}
\end{figure*}

\begin{figure*}[t!]
(a)\includegraphics[scale=0.3]{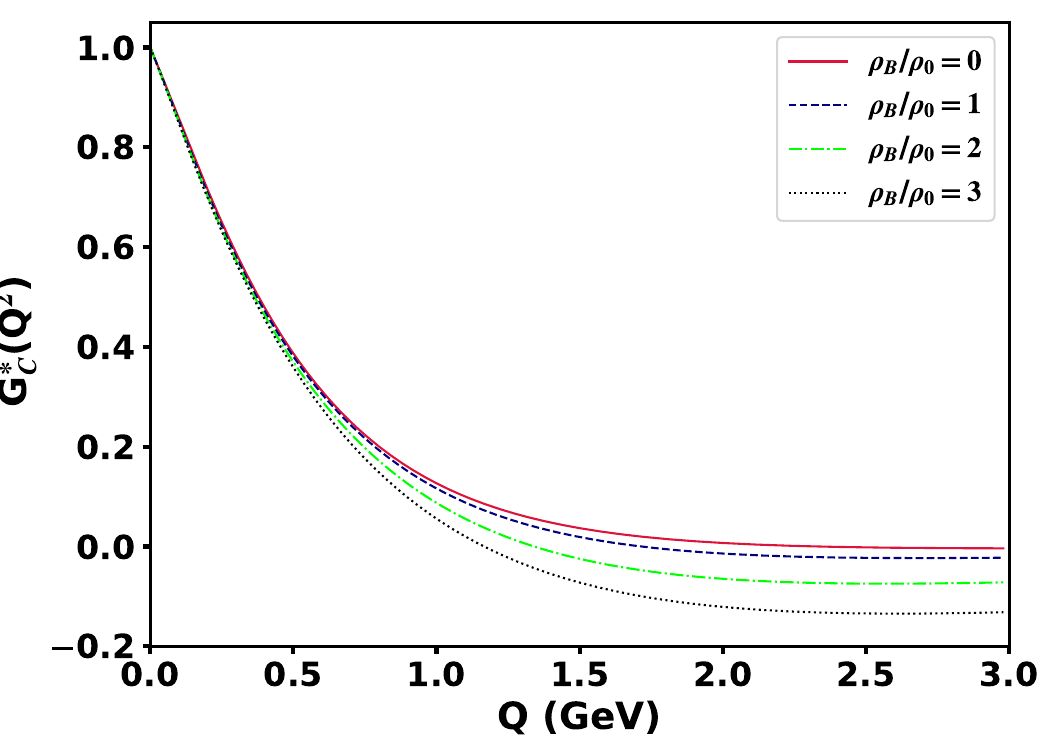}
(b)\includegraphics[scale=0.3]{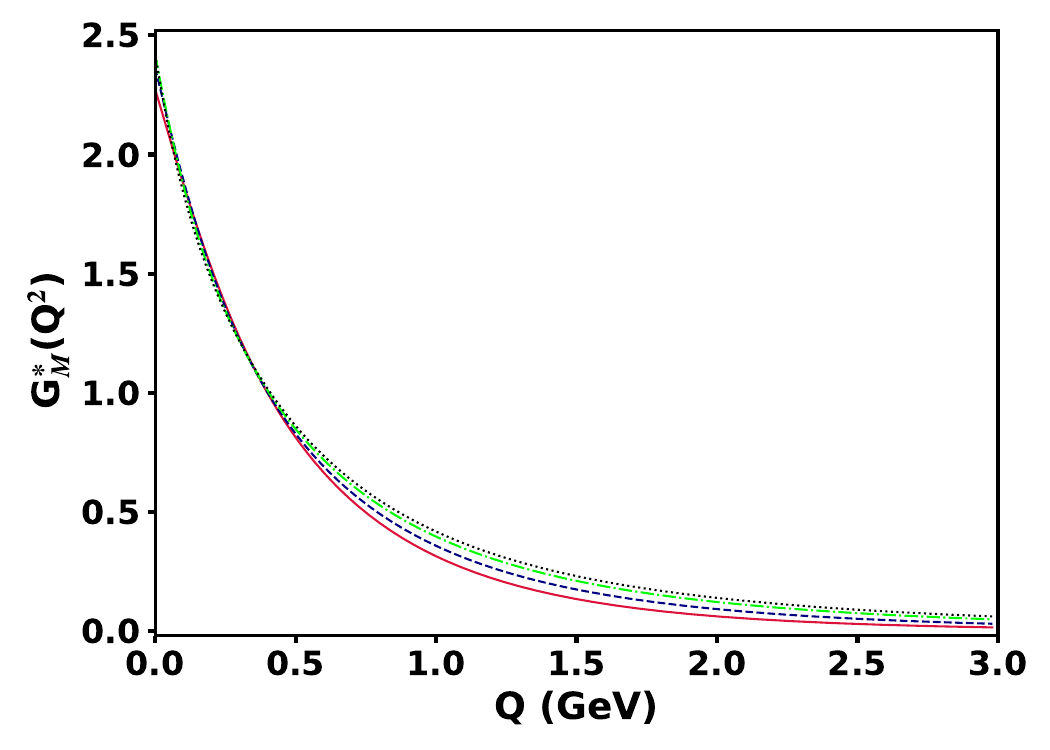}
(c)\includegraphics[scale=0.3]{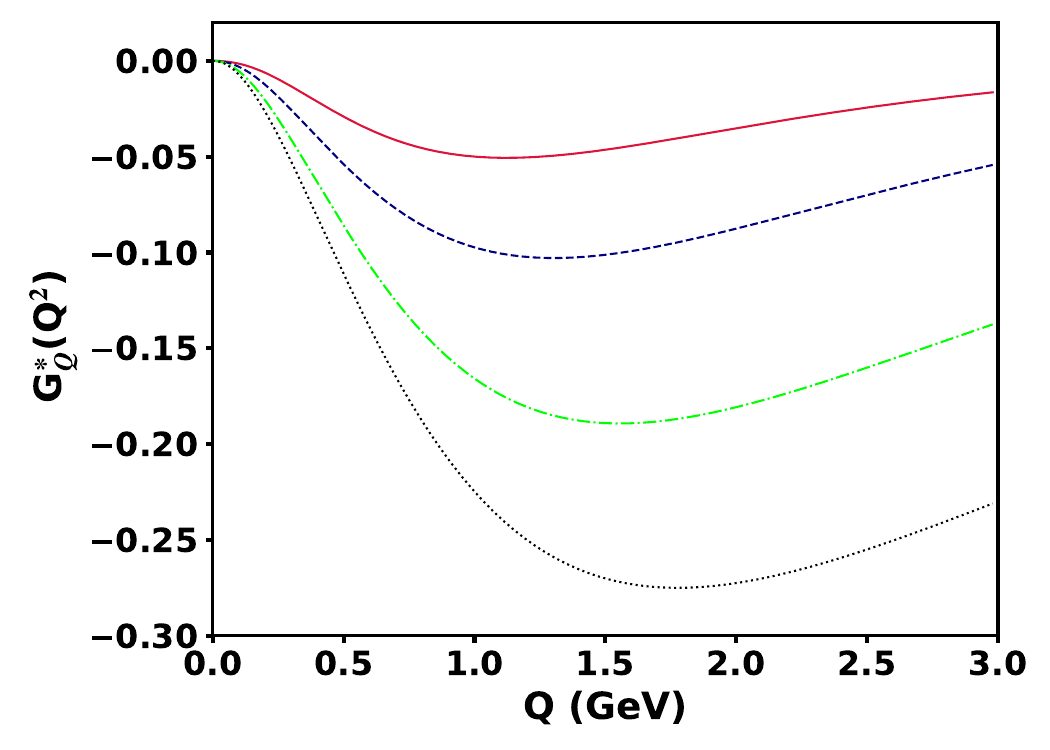}
\caption{ Variation of the medium modified (a) charge \((G^{\ast}_{C}(Q^2))\), (b) magnetic \((G^{\ast}_{M}(Q^2))\), and (c) quadrupole \((G^{\ast}_{Q}(Q^2))\) form factors with momentum transfer Q (GeV) shown at baryonic density ratios ($\rho_B/\rho_0$) varying from 0 to 3 at an interval of 1.}\label{plot2}
\end{figure*}

In Fig.~\ref{plot1}, we have plotted the medium modified charge \((G^{\ast}_{C}(Q^2))\), magnetic \((G^{\ast}_{M}(Q^2))\), and quadrupole \((G^{\ast}_{Q}(Q^2))\) form factors versus the momentum transfer Q (GeV) at baryonic density ratio varying from 0 to 1 (in intervals of 0.25). We have also compared our results with the available lattice simulation data. We can observe the rapid decline of \(G^{\ast}_{C}(Q^2), G^{\ast}_{M}(Q^2)\) and \(G^{\ast}_{Q}(Q^2)\), with increasing value of Q. At lower values of baryonic density ratio \(\rho_{B}/\rho_{0}\), the variation in all the form factors with respect to Q remain minimal, indicating weaker medium effects. The lattice QCD results show qualitative agreement with the theoretical curves, particularly at low Q, though noticeable deviations and uncertainties appear at higher momentum transfer values. To study the effect of medium at higher baryonic ratios, Fig.~\ref{plot2} illustrates the medium modified charge, magnetic, and quadrupole form factors versus Q (GeV) at baryonic density ratio \(\rho_{B}/\rho_{0}\) from 0 to 3 (in intervals of 1). In the subplot (a) of Figs.~\ref{plot1} and~\ref{plot2}, we notice that the medium modified charge form factor \(G^{\ast}_{C}(Q^2)\) exhibits a significant reduction with increasing baryonic density at high Q. This indicates that the internal charge distribution of the \(\rho\) meson becomes more significantly altered in a denser symmetric nuclear medium. Furthermore, \(G^{\ast}_{C}(Q^2)\)  crosses zero at progressively lower values of Q as the value of the baryonic density ratio rises. These zero crossings are found at Q$=$2.32, 2.15, 1.99, 1.85, 1.71, 1.34, and 1.16 GeV for the baryonic density ratio value \(\rho_{B}/\rho_{0}=0, 0.25, 0.5, 0.75, 1, 2 \text{ and }3\), respectively. This shift reflects the influence of the nuclear medium on the structure and behavior of the \(\rho\) mesons. In subplot (b) of Figs.~\ref{plot1} and~\ref{plot2}, we observe that the medium modified magnetic form factor \(G^{\ast}_{M}(Q^2)\) exhibits a decreasing trend similar to that of behavior of  \(G^{\ast}_{C}(Q^2)\) with the increasing value of Q (GeV). In contrast, the magnitude of \(G^{\ast}_{M}(Q^2)\) increases as the baryonic density ratio \(\rho_{B}/\rho_{0}\) rises from 0 to 3. This upward shift suggests that the nuclear medium enhances the magnetic characteristics of the nucleon, particularly at higher densities. The medium modified quadrupole form factor \(G^{\ast}_{Q}(Q^2)\) is plotted as a variation of momentum transfer Q, with increasing values of baryonic density ratio \(\rho_{B}/\rho_{0}\) in subplot (c) of Figs.~\ref{plot1} and~\ref{plot2}. Starting from zero, the curves drop into negative values and gradually flattens out at higher Q.  This trend is observed consistently across all baryonic density ratios. As the baryonic density ratio increases, the minimum value of \(G^{\ast}_{Q}(Q^2)\) becomes more negative, indicating a strong interaction of the \(\rho\) mesons and the symmetric nuclear medium.

\begin{figure*}[t!]
(a)\includegraphics[scale=0.3]{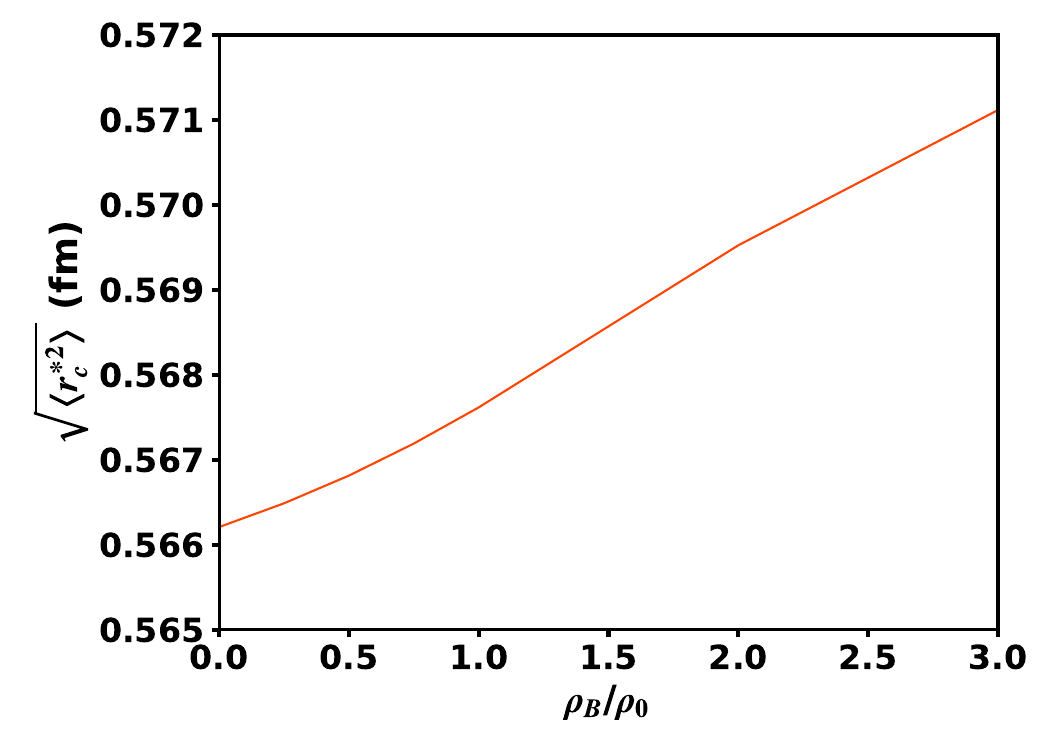}
(b)\includegraphics[scale=0.3]{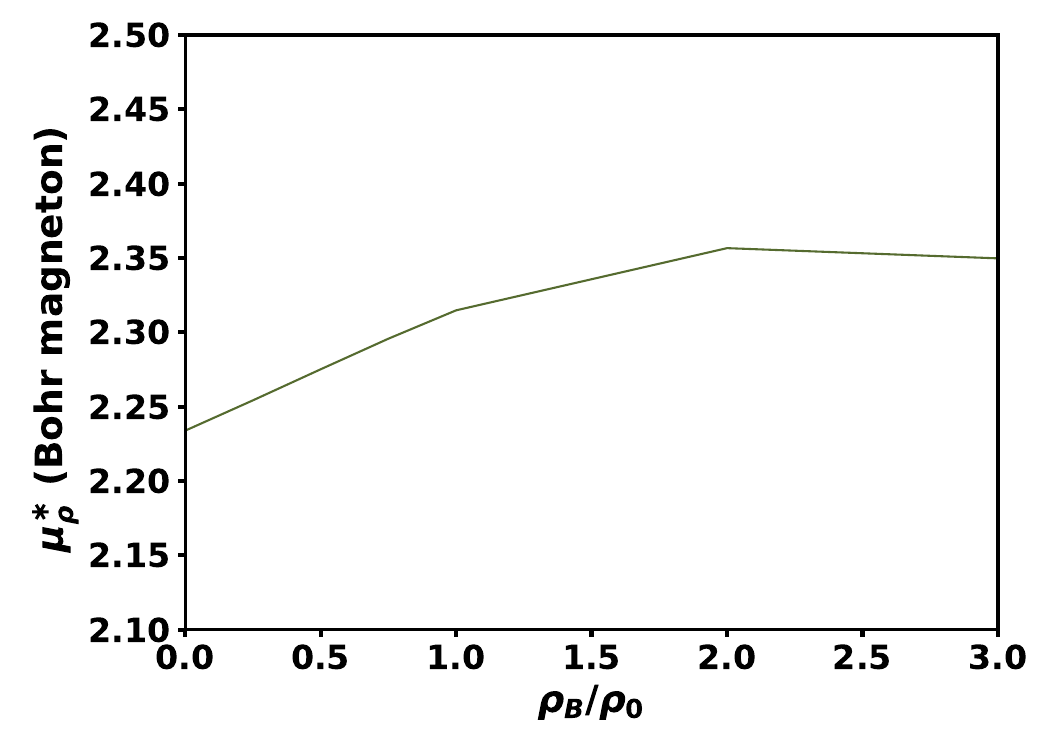}
(c)\includegraphics[scale=0.3]{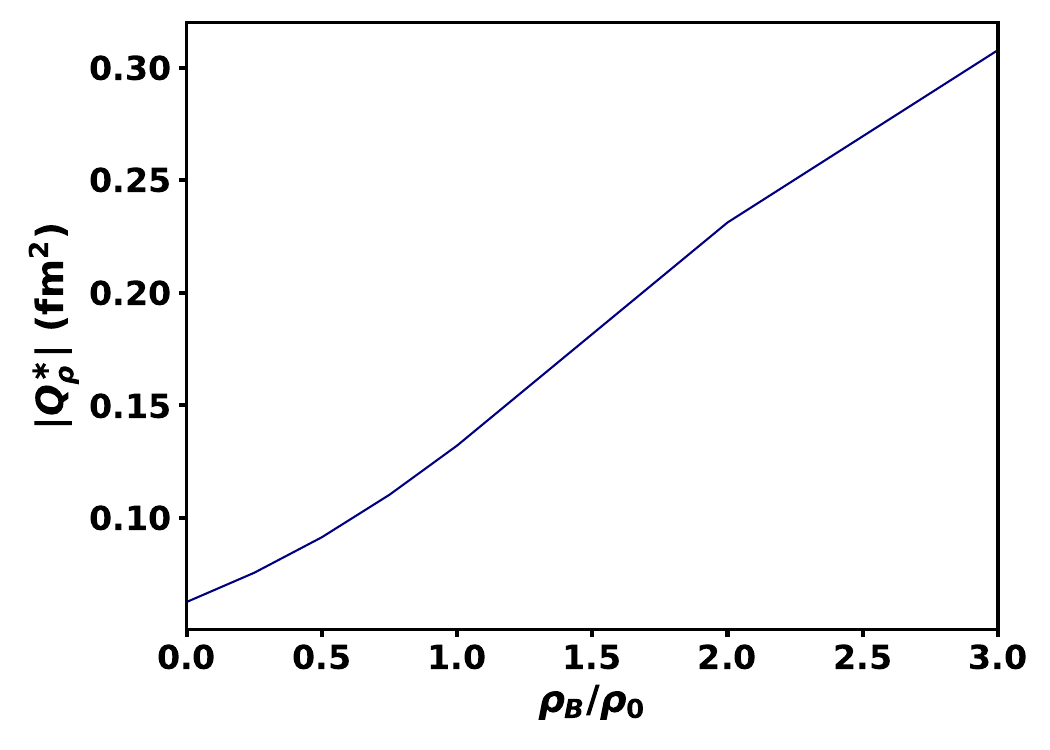}
\caption{ The medium modified (a) charge radii (\(\sqrt{\langle r_c^{\ast 2}\rangle}\)), (b) magnetic moment (\(\mu^{\ast}_p\)), and (c) quadrupole moment (\( \left|Q^{\ast}_p \right|\)) are plotted as functions of the baryonic density ratio ($\rho_B/\rho_0$). }\label{plot3}
\end{figure*}

The variation of charge radii (\(\sqrt{\langle r_c^{\ast 2}\rangle}\)), magnetic moment (\(\mu^{\ast}_{\rho}\)), and quadrupole moment (\(Q^{\ast}_{\rho}\)) with an increase in baryonic density ratio \(\rho_{B}/\rho_{0}\) (from 0 to 3) is shown in Fig.~\ref{plot3}. As shown in subplot (a) of Fig.~\ref{plot3}, the value of modified charge radii of \(\rho\) meson is found to increase progressively with baryonic density ratio. This behavior in the symmetric nuclear medium, considered alongside the decreasing in-medium mass of the \(\rho\) meson with increasing baryonic density ratio, suggests medium-induced delocalization, where the \(\rho\) meson appears to be less tightly bound. In subplot (b) of Fig.~\ref{plot3}, the medium modified magnetic moment (\(\mu^{\ast}_{\rho}\)) shows an increase with increasing baryonic density ratio  \(\rho_{B}/\rho_{0}\) (upto \(\rho_{B}/\rho_{0} \rightarrow 2\)), then slightly decreases. This trend suggests that the magnetic properties of the meson are significantly influenced by the surrounding nuclear medium. This behavior indicates that the symmetric nuclear medium enhances the \(\rho\) meson's magnetic response at moderate densities, but this effect weakens at higher densities due to partial restoration of symmetries. Subplot (c) of Fig.~\ref{plot3} displays a steady increase in the absolute value of the quadrupole moment (\(Q^{\ast}_{\rho}\)) of \(\rho\) meson with rising baryonic density ratio  \(\rho_{B}/\rho_{0}\). This trend suggests that the symmetric nuclear medium induces greater deformation in the meson’s internal charge distribution, reflecting increased spatial distortion at higher baryonic densities. 

Table~\ref{table1} presents a quantitative comparison of our calculated values for \(\sqrt{\langle r_c^{\ast 2}\rangle}, \mu^{\ast}_{\rho}, \text{ and } Q^{\ast}_{\rho}\) with the available results from other theoretical approaches in Refs.~\cite{Choi:2004ww, Bhagwat:2006pu, Hernandez-Pinto:2024kwg, Luan:2015goa, Shultz:2015pfa, Owen:2015gva, Tanisha:2025qda}. Our results are consistent with those obtained in free space, demonstrating compatibility with existing models.

\begin{table*}[t]
\renewcommand{\arraystretch}{1.5}
    \centering
    \begin{tabular}{ | c |c|c|c|c|}
    \hline\hline
          &$\sqrt{\langle r^{\ast 2}_c \rangle}$  & $ \mu^{\ast}_{\rho}$   & $ Q^{\ast}_{\rho} $   \\\hline
          Our (Vacuum) &0.5662 & 2.23 & -0.0626   \\
         Ref. \cite{deMelo:2018hfw}&0.51 & 2.20 & -0.059   \\
         Ref. \cite{Hutauruk:2025bjd}&0.67 & 2.53 & -0.057   \\
        Ref. \cite{Bhagwat:2006pu} & 0.73 & 2.01 & -0.026 \\
         Ref. \cite{Owen:2015gva} & 0.82& 2.07 & -0.045    \\
         Ref.  \cite{Shultz:2015pfa}&0.82 & 2.48 & -0.070   \\
         Ref. \cite{Hernandez-Pinto:2024kwg}& 0.56& 0.75 & \dots   \\
         Ref. \cite{Luan:2015goa}&1.12 & 2.54 & \dots    \\
           Ref. \cite{Choi:2004ww}&0.52 & 1.92 & -0.028    \\
          Ref. \cite{Tanisha:2025qda}&0.95 & 2.19 & -0.023   \\  
          \hline
            Our ($\rho_B=0.25\rho_0$) & 0.5664& 2.2542 &-0.0756    \\
            Ref. \cite{deMelo:2018hfw}&0.60 & 2.18 & -0.071   \\
            \hline
            Our ($\rho_B=0.50\rho_0$) &0.5668 &2.2751  &-0.0914    \\
            Ref. \cite{deMelo:2018hfw}&0.74 & 2.18 & -0.087  \\
            Ref. \cite{Hutauruk:2025bjd}&0.71 & 2.41 & -0.065   \\
            \hline
             Our ($\rho_B=0.75\rho_0$) &0.5671 &2.2956  & -0.1102    \\
            \hline
             Our ($\rho_B=\rho_0$) &0.5676 & 2.3147 & -0.1321 \\
            Ref. \cite{Hutauruk:2025bjd}&0.75 & 2.33 & -0.071  \\
            \hline
             Our ($\rho_B=2 \rho_0$) &0.5695 &2.3566  & -0.2312   \\
            Ref. \cite{Hutauruk:2025bjd}&0.79 & 2.26 & -0.080   \\
            \hline
             Our ($\rho_B=3 \rho_0$) &0.5711 &2.3497  & -0.3078   \\
    \hline \hline    
    \end{tabular}
\caption{The charge radii $\sqrt{\langle r^{\ast2}_c \rangle}$ (fm), magnetic moment $\mu^{\ast}_{\rho}$ (in units of Bohr magneton), and quadrupole moment $Q^{\ast}_{\rho}$ (fm$^2$) of $\rho$ meson have been compared with available data \cite{deMelo:2018hfw,Choi:2004ww, Bhagwat:2006pu, Hernandez-Pinto:2024kwg, Luan:2015goa, Shultz:2015pfa, Owen:2015gva, Tanisha:2025qda}.}
    \label{table1}
\end{table*}


 \section{Conclusions}
\label{Sec7}
In this study, we systematically examined the impact of the symmetric nuclear medium on the properties of the vector \(\rho\) meson using the combined framework of LFQM and the CQMF model. For this purpose, the in-medium modified quark masses are first obtained from the CQMF model and then utilized as input in the LFQM.

\par To begin with, we obtained the vacuum mass and the weak decay constant of the \(\rho\) meson by fitting the ground-state mass spectra to determine the \(\beta\) parameter through the variational principle. The obtained vacuum results are found to be consistent with both the experimental data reported by the PDG and previous theoretical predictions. The study then investigates the behavior of the \(\rho\) meson in a nuclear medium, examining its modified mass, weak decay constant, distribution amplitudes (DAs), decay width, Mellin moments, and form factors (charge, magnetic, and quadrupole), along with the associated charge radii, magnetic moment, and quadrupole moment across different baryon densities. It was observed that the effective mass of the \(\rho\) meson decreases at lower baryon densities, then gradually increases when the baryon density \(\rho_B \ge \rho_0\), which can be attributed to the inclusion of the hyperfine potential in the calculations. The in-medium decay constant \(f^*_\rho\) and the normalized decay constant \(f^*_\rho/f_\rho\) both exhibit a gradual decline as the baryon density increases. However, the longitudinal decay constant \(f^{\parallel *}_\rho\) exhibited only a slight variation, while the transverse decay constant \(f^{\perp *}_\rho\) showed a more pronounced reduction in magnitude as the density increased. Furthermore, the decay width of the \(\rho\) meson was analyzed both in vacuum (\(\Gamma_\rho\)) and at different baryon densities (\(\Gamma^*_\rho\)).  It was observed that \(\Gamma^*_\rho\) increased up to \(\rho_B=\rho_0\) before decreasing as the density continued to increase. Using the medium-modified decay width, the branching ratio and associated lifetime of the \(\rho\) meson in the medium were found to undergo only slight variations as the baryon density increased.

\par To gain further insight into the internal dynamics of the \(\rho\) meson, the in-medium behavior of longitudinal \(\phi^{\parallel *}_\rho\) and transverse \(\phi^{\perp*}_\rho\) DAs was also examined up to \(\rho_B=3\rho_0\). Both DAs were observed to decrease with an increase in density for \(x\) values ranging from \(0.225\) to \(0.775\), while an increase was observed for all other values \(x\). Furthermore, the evolved DAs were calculated using the Efremov–Radyushkin–Brodsky–Lepage (ERBL) evolution equations, and the obtained results showed a reasonable consistency and a matching trend with the lattice simulation and ILM results. The Mellin moments and Gegenbauer moments were evaluated up to \(n=8\) for both vacuum and medium DAs up to \(\rho_B=3\rho_0\). The vacuum results were found to be in good agreement with previous theoretical studies. In addition, the Mellin, inverse Mellin, and Gegenbauer moments exhibited a gradual increase with increasing baryon density.

\par We have also analyzed the influence of a symmetric nuclear medium on the charge \((G^{\ast}_{C}(Q^2))\), magnetic \((G^{\ast}_{M}(Q^2))\), and quadrupole \((G^{\ast}_{Q}(Q^2))\) form factors of spin-1 \(\rho\) mesons. The in-medium form factors \((G^{\ast}_{C}(Q^2))\), \((G^{\ast}_{M}(Q^2))\), and \((G^{\ast}_{Q}(Q^2))\) exhibit a decreasing trend with increasing momentum transfer \(Q\) (in GeV), which aligns well with the behavior reported in lattice simulation data. With an increase in the baryonic density ratio \(\rho_B/\rho_0\), the in-medium charge form factor \((G^{\ast}_{C}(Q^2))\) of the \(\rho\) meson decreases, while the magnetic \((G^{\ast}_{M}(Q^2))\) and quadrupole \((G^{\ast}_{Q}(Q^2))\) form factors show an opposite trend. Furthermore, the impact of increasing baryonic density ratio \(\rho_B/\rho_0\) (up to \(\rho_B = 3\rho_0\)) on the medium modified charge radius \((\sqrt{\langle r_c^{\ast 2}\rangle})\), magnetic moment \((\mu^{\ast}_{\rho})\), and quadrupole moment \((|Q^{\ast}_{\rho}|)\) of the \(\rho\) meson has been investigated. The results indicate that all three observables increase with rising baryonic density ratio, indicating an enhancement of the \(\rho\) meson’s internal structure within a denser nuclear medium. A comparison with other theoretical approaches shows a consistent qualitative agreement, supporting the credibility and physical relevance of the present investigation.

\par This work advances our understanding of vector meson behavior in dense nuclear environments and establishes a consistent theoretical framework for studying their modifications in medium. The present study offers valuable insights into how QCD dynamics evolve as baryon density increases, suggesting potential indications of partial chiral symmetry restoration within nuclear matter. This provides a solid foundation for examining in-medium observables, such as mass shifts and spectral functions, which are currently being studied in dilepton at NA60 \cite{NA60:2006ymb}, CERES \cite{CERES:2006wcq}, HADES \cite{HADES:2011nqx}, and the upcoming CBM-FAIR program \cite{CBM:2016kpk}. Furthermore, the developed framework can be employed in future analyses of other non-perturbative QCD phenomena, such as electromagnetic interactions, parton-level distributions, and temperature-related effects, and it can also be expanded to investigate other light and heavy vector mesons or higher-twist contributions to distribution amplitudes.

\section{Acknowledgement}
H.D. would like to thank  the Science and Engineering Research Board, Anusandhan-National Research Foundation, Government of India under the scheme SERB-POWER Fellowship (Ref No. SPF/2023/000116) for financial support. A. K. sincerely acknowledge Anusandhan National
Research Foundation (ANRF), Government of India for
funding of the research project under the Science and
Engineering Research Board-Core Research Grant
(SERB-CRG) scheme (File No. CRG/2023/000557).

\bibliographystyle{apsrev}  
\bibliography{ref} 
\end{document}